\documentclass[aps,pre,twocolumn,floatfix]{revtex4}
\usepackage{graphicx}
\usepackage{amsmath}
\usepackage{soul}
\usepackage{standalone}

\newcommand{\gammaeff}{\gamma_{\text{eff}}}

\newlength\figurewidth
\usepackage{color}

\begin{document}

\title{Active microrheology in corrugated channels: comparison of thermal and colloidal baths}
\date{\today}
\def\affPME{\affiliation{%
Helmholtz Institute Erlangen-N\"urnberg for Renewable Energy (IEK-11), Forschungszentrum J\"ulich, F\"urther Stra$\beta$e 248,
90429 N\"urnberg, Germany
  }}
  \def\affPMS{\affiliation{%
  Max Planck Institute for Intelligent Systems,   Heisenbergstr.\ 3, 70569 Stuttgart,   Germany}}
\def\affPMSU{\affiliation{IV Institute for Theoretical Physics, University of Stuttgart, Pfaffenwaldring 57,\! 70569 Stuttgart, Germany
  }}

\def\affAP{\affiliation{%
  Departamento de Qu\'{\i}mica y F\'{\i}sica, Universidad de Almer\'ia,
  04.120 Almería, Spain}}

\def\affIPCH{\affiliation{Centre  Europ\'een  de  Calcul  Atomique  et  Mol\'eculaire (CECAM), Ecole  Polytechnique  F\'ed\'erale  de  Lausanne (EPFL),  Batochimie,  Avenue  Forel  2,  1015  Lausanne,  Switzerland}}

\def\affIPBCNN{\affiliation{UBICS  University  of  Barcelona  Institute  of  Complex  Systems,  Mart\'{\i}  i  Franqu\`es  1,  E08028  Barcelona,  Spain}
}

\def\affIPBCN{\affiliation{Departament de Fisica de la Materia Condensada, Universitat de Barcelona, Mart\'{\i}  i Franqu\`es 1, 08028 Barcelona, Spain}}
  
\author{P.~Malgaretti}\affPMS\affPMSU\affPME
\author{A.~M.~Puertas}\affAP
\author{I.~Pagonabarraga}\affIPCH\affIPBCN\affIPBCNN

\begin{abstract}
\noindent HYPOTHESIS\\
The dynamics of colloidal suspension confined within porous materials strongly differs from that in the bulk.
In particular, within porous materials, the presence of boundaries with complex shapes entangles the longitudinal and transverse degrees of freedom inducing a coupling between the transport of the suspension and the density inhomogeneities induced by the walls.\\
\noindent METHOD\\
Colloidal suspension confined within model porous media are characterized by means of active microrheology where a net force is applied on a single colloid (tracer particle) whose transport properties are then studied. The trajectories provided by active microrheology are exploited to determine the local transport coefficients. In order to asses the role of the colloid-colloid interactions we compare the case of a tracer embedded in a colloidal suspension to the case of a tracer suspended in an ideal bath.\\
\noindent FINDING\\
Our results show that the friction coefficient increases and the passage time distribution widens upon increasing the corrugation of the channel. These features  are obtained for a tracer suspended in a (thermalized) colloidal bath as well as for the case of an ideal thermal bath. These results highlight the relevance of the confinement on the transport and show a mild dependence on the colloidal/thermal bath.
Finally, we rationalize our numerical results with 
 a semi-analytical model. Interestingly, the predictions of the model are quantitatively reliable for mild external forces, hence providing a reliable tool for predicting the transport across porous materials.
\end{abstract} 

\maketitle


\section{Introduction}
Microrheology has been introduced recently to study the microscopic diffusion of particles in complex environments \cite{Furst2017}. According to this technique, a tracer colloidal particle is introduced in the complex bath, and its diffusion is monitored, typically by microscopy. In passive microrheology, the tracer is allowed to diffuse and explore different regions \cite{Mason1995,Cicuta2007,Wilson2009}, whereas in active microrheology, an external force is applied to the tracer, to make it travel long distances, allowing for the calculation of a microscopic friction coefficient and structural heterogeneities \cite{Habdas2004,Squires2005,Gazuz2009,Puertas2014,Zia2018}. Theoretical models have been developed for bulk systems, relating the tracer dynamics with the properties of the bath, both for low densities~\cite{Squires2005,DePuit2011,Chu2019}, and dense systems \cite{Gazuz2009,Gruber2016,Su2017}. Moreover, in the case of very strong pulling forces, non-linear response has been found for both bulk~\cite{Benichou2014} as well as for channels with constant sections~\cite{Illien2014,Benichou2016,Benichou2018} . Microrheology has been applied to simple systems, where the models have been developed, but also to more complex systems beyond colloids, such as polymer or fiber solutions \cite{Gisler1999,Oppong2005}, soaps \cite{Prasad2009}, and particularly biological environments, such as mucus \cite{Weigand2017}, or the interior of living cells \cite{Nishizawa2017}.

\begin{figure}
\includegraphics[scale=1]{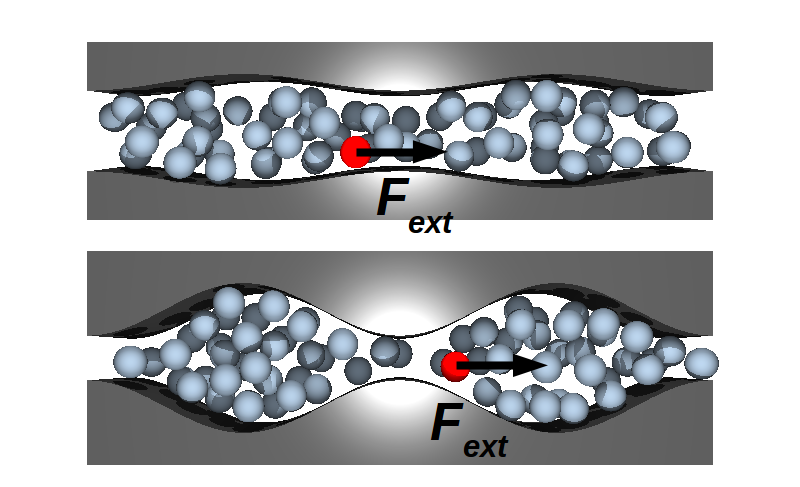}
\caption {Snapshot of the systems with corrugation amplitudes of $A=0.51a$
(upper panel) and $A=1.62a$ (lower panel). In both cases, $L_z=L_x=6\,a$ and $L_y=40\,a$. The red particle is the tracer and the arrows indicate the direction of the force. \label{snapshots}}
\end{figure}
Accordingly, active microrheology seems a promising technique to study the transport of macromolecules, polymers or colloids suspended in complex fluids embedded in porous materials. Indeed, the spatial and time resolution probed by active microrheology can provide insight into the dependence of the transport on the local structural (density) modulations induced by the confining walls, and disentangle it from the effects of the complex fluid. The coupling of both effects can induce important modifications in the transport mechanism, such as non-isotropic diffusion \cite{Marconi2015,Puertas2018}, that eventually may be exploited to tune the transport of the tracer.


The simplest model porous material is a one-dimensional channel as shown in Fig.\ref{snapshots}. When the confining walls are non-planar, the confinement induces structural (density) modulations, that affect also the transport properties along the principal channel direction. The solution of the general problem thus requires the solution of the three-dimensional diffusion equation, or the Fokker-Plank equation at the microscopic level, with the geometric constraints imposed by the channel. Alternatively, the problem can be studied effectively by the Fick-Jacobs approximation, that assumes that the equilibration in the transversal direction is much faster than the diffusion along the channel direction \cite{Jacobs1967,Zwanzig1992}. This results again in a one-dimensional problem, where the external potentials (or confining walls) can be introduced. This description was improved by Zwanzig~\cite{Zwanzig1992}, who introduced a space-dependent diffusion coefficient, and further developed in different works  \cite{Reguera2001,Kalinay2006,Dagdug2012_2,dagdug2015,Kalinay2018}.


In this contribution, we exploit active microrheology to study the transport of colloids suspended in complex fluids embedded in  porous materials. In particular, a colloidal suspension of hard particles containing a single tracer bead is confined in a varying-section channel. The tracer particle is subject to an external force, i.e. we implement the active microrheology protocol in a corrugated channel. We perform numerical simulations in which all particles have the same size and undergo Langevin dynamics. The average channel width is three particle diameters, and the narrowest point in the channel, channel neck, has been varied from five times the particle radius to three times. For small external forces, this implies a strong increase of the friction coefficient and mean first passage times, whereas for large forces, the effect of the corrugation amplitude is almost unnoticed.

In order to disentangle the dependence of the transport of the tracer on the properties of the complex fluid (i.e. the colloidal suspension) from the dependence on the shape of the channel,  we compare our results against the case of a tracer suspended in an ideal thermal bath (Newtonian solvent) embedded in the same channel. The numerical results are also compared with a theoretical model that is based on the Fick-Jacobs approximation~\cite{Jacobs1967,Zwanzig1992,Reguera2001,Malgaretti2013}. Within this model, the geometrical confinement is accounted for by an effective entropic force; the agreement of the model with the numerical data is quantitative for the isolated tracer and qualitative when the colloidal bath is included. This approach has provided insight for quite diverse confined systems including colloids \cite{Martens2011,dagdug2015}, flow of (charged) fluids~\cite{Hanggi2013,Malgaretti2014,Chinappi2018,Malgaretti2019JCP,Kalinay2020}, polymers~\cite{Bianco2016,Malgaretti2019,Carusela2021}, rigid rods~\cite{Malgaretti2021},  chemical reactions~\cite{Ledesma-Duran2016}, and pattern-forming systems~\cite{Chacon-Acosta2020}.

Our results show that active microrheology is a useful tool to characterize the transport of colloids suspended in confined complex fluids. In particular, with our technique we have been able to determine the local tracer position distribution and velocity. Moreover, the good match between our numerical results and the Fick-Jacobs model confirms the validity of the model to predict the local transport properties and to relate them to the macroscopic transport. Interestingly, our data show that some features, such as the tracer density profile or the Mean First Passage Time (MFPT) are robust with respect to the properties of the fluid the tracer is suspended in and their qualitative behavior can be predicted by our Fick-Jacobs model despite the bath is neglected. 


\section{Numerical model}

In the simulations, we consider a colloidal system of moderate density confined in a corrugated channel, with a tracer particle. All particles, including the tracer, have the same radius, $a$, and undergo Brownian motion. In addition to this random motion, an external force, parallel to the channel direction, acts on the tracer~\cite{Puertas2018}. 
The confining walls are defined by the corrugated surfaces:

\begin{equation}
h(y)=\pm \left(\frac{L_z}{2} - A \cos \frac{2\pi y}{\lambda} \right) 
\end{equation}

\noindent where $A$ and $\lambda$ are the corrugation amplitude and wavelength, respectively, and $L_z$ sets the mean separation of the walls, which we fix equal to $3$ particle diameters, i.e. $L_z=6a$. 
For later use we introduce the entropy barrier  which for finite-size particles is defined as \cite{Malgaretti2016}:

\begin{equation}
\Delta S = \ln \left[\frac{h_{max}-a}{h_{min}-a}\right]
\label{eq:DS}
\end{equation}

\noindent with $h_{max}$ and $h_{min}$ the maximum and minimum channel widths, respectively. 
In all cases, the corrugation wavelength is $\lambda=20a$, and the volume fraction of the colloidal bath is $\phi=0.20$. The system has periodic boundary conditions in the $XY$ and $XZ$ planes. The snapshots in Fig. \ref{snapshots} show the systems with the lowest ($A=0.5\,\Delta S=0.33$) and highest ($A=1.2\,\Delta S=1.2$) corrugation amplitude. 

In order to disentangle the effects of the bath and confinement, simulations without the colloidal bath have also been performed. In this case only the tracer confined between the walls is simulated, undergoing Brownian motion (due to the solvent, or thermal bath) and pulled by the external force.

Previous works of active microrheology in bulk or in channels have shown that the effective friction experienced by the tracer develops a plateau at low forces, what allows the definition of a linear regime \cite{Squires2005,Carpen2005,DePuit2011,Puertas2014,Puertas2018}. Upon increasing the external force the effective friction decreases, i.e. the system experiences a so called, force-thinning regime. Finally, for large forces the friction reaches a second plateau, at a lower value than the low-force plateau. Therefore, we have selected three values of the external force, 
$\beta F_{ext} \lambda=10$, $100$ and $1000$, corresponding to the low-P\'eclet plateau, force thinning, and high-P\'eclet plateau, respectively. In particular cases, other values of the force have been simulated.
Further details on the simulation method can be found in the Supplemental Material.
\section{Theoretical model}

For a theoretical description of the system under consideration, we have used a model based on the Fick-Jacobs approximation of the transport properties of a channel,  described in Ref.\cite{Malgaretti2016}. 
All quantities in the theoretical model are intended in SI units.
Briefly, after integration along the transverse direction, the time evolution of the distribution of the tracer position along the channel axis, $P(y,t)$, is given by:

\begin{equation}
\frac{\partial \rho_y(y,t)}{\partial t} = \frac{\partial}{\partial y} \left[ \beta D(y) \rho_y (y,t) \frac{\partial {\cal A}(y)}{\partial y} + D(y) \frac{\partial \rho_y (y,t)}{\partial y} \right]
\label{eq:FJ}
\end{equation}
where $D(y)$ is the position-dependent diffusion coefficient and  ${\cal A}(y)$ is the potential of mean force~\cite{Marconi2015,Puertas2018}. The steady-state solution of this equation reads:
\begin{align}
    \rho_y (y)=e^{-\beta \mathcal{A}(y)}\left[-J\int_{-\frac{\lambda}{2}}^y \dfrac{e^{\beta \mathcal{A}(y')}}{D(y)} dy' +\Pi\right]
    \label{eq:sol-FJ}
\end{align}
with
\begin{align}
    \Pi &=-J\dfrac{\displaystyle\int_{-\frac{\lambda}{2}}^\frac{\lambda}{2} \dfrac{e^{\beta \mathcal{A}(y')}}{D(y')}dy'}{e^{-\beta \Delta\mathcal{A}}-1}=-J \Pi_0\\
    J&=-\left[\int_{-\frac{\lambda}{2}}^\frac{\lambda}{2}e^{-\beta \mathcal{A}(y)}\left[\int_{-\frac{\lambda}{2}}^y \dfrac{e^{\beta \mathcal{A}(y')}}{D(y')} dy' +\Pi_0\right]dy\right]^{-1}\label{eq:flux}
\end{align}
where $\Delta\mathcal{A}=\mathcal{A}(-\lambda/2)-\mathcal{A}(\lambda/2)$ and $\Pi$ and $J$ are determined by the periodic boundary condition and by the normalization
\begin{align}
    \int_{-\frac{\lambda}{2}}^\frac{\lambda}{2}P(y)dy=1
\end{align}
For the case of a single tracer suspended in an ideal gas, ${\cal A}(y)$ encodes only the confining forces and the external pulling force, $F_{ext}$: 
\begin{align}
 {\cal A}(y)=-k_{\text{B}}T\ln\left[\frac{h(y)}{h_0}\right]-F_{ext}y
 \label{eq:DA}
\end{align}
whereas for interacting systems the functional form of ${\cal A}(y)$ is more involved~\cite{Puertas2018}. 
In the following we assume a constant diffusion coefficient along the channel, i.e.
\begin{align}
 D(y)=D_0
\end{align}
where $D_0$ is the bulk diffusion coefficient of the colloid.

We use Eq.~\eqref{eq:sol-FJ} to determine the ratio between the probability of finding the tracer in the two halves of the channel $\rho_L/ \rho_R$ which, in the limit of small forces, reduces to (see Suppl. Mat.):
\begin{equation}
\frac{\rho_L}{\rho_R} = 1 + \mathcal{H} \beta \Delta S F_{ext}\lambda\,.
\label{eq:rho-ratio}
\end{equation}
where $\mathcal{H}$ is a prefactor whose value depends on the specific shape of the channel. 
\noindent We remark that in the expansion the product $\beta \Delta S F_{ext}\lambda$ naturally appears as the small parameter controlling the magnitude of higher order terms. This underlines that in the small-force regime $\Delta S$ and $\beta F_{ext}\lambda$ have a similar effect on the asymmetry in the distribution of the tracer.

\subsection{Mean first passage time (MFPT)}

A quantity that is often of interest in diffusion processes ranging from neuronal signaling to nuclear waste container~\cite{Rednerb} is the Mean First Passage Time (MFPT) i.e., the mean time that a particle takes to reach a certain location for the first time~\cite{Redner}. 
Starting from Eq.~\eqref{eq:FJ}, it is possible to derive an equation for the MFPT~\cite{Gardiner_book}, $t_1(0)$, for a particle over a period of the channel $\lambda$ (see Suppl. Mat.):
\begin{flalign}\label{eq:MFPT_tot}
&t_{1}(0) =\frac{-\lambda^{2}}{D\left[(\beta F_\text{ext}\lambda)^{2}-(\beta\Delta{\cal F}\lambda)^{2}\right]^{2}}\times\nonumber &&\\
&\times\Bigg[e^{-\beta F_\text{ext}\lambda}((\beta\Delta{\cal F}\lambda)^{2}-(\beta F_\text{ext}\lambda)^{2})+\\
&+\left(\beta F_\text{ext}\lambda\right)^{2}-\left(\beta F_\text{ext}\lambda\right)^{3}+3\left(\beta\Delta{\cal F}\lambda\right)^{2}+\beta F_\text{ext}\lambda\left(\beta\Delta{\cal F}\lambda\right)^{2}\nonumber\\
 & +\dfrac{2\beta(F_\text{ext}\lambda-\Delta{\cal F}\lambda)\Delta{\cal F}\lambda-2e^{\beta \lambda\Delta{\cal F}}\Delta{\cal F}\lambda(F_\text{ext}\lambda+\Delta{\cal F}\lambda)}{e^{\frac{1}{2}\beta \lambda(F_\text{ext}+\Delta{\cal F})}}\Bigg]\nonumber\,.
\end{flalign}
where we used the effective force
\begin{equation}\label{eq:DF}
{\cal F}(y) =  \left\{ \begin{array}{lll}
{\cal F}_1=F_{ext} + \Delta \mathcal{F} & \hspace{1cm} & y < 0 \\ 
{\cal F}_2=F_{ext} - \Delta \mathcal{F} & \hspace{1cm} & y > 0
\end{array} \right.
\end{equation}
derived from the free energy barrier 
\begin{align}
\Delta \mathcal{F}=-\frac{\mathcal{A}(0)-\mathcal{A}(\lambda/2)}{\lambda/2}.
\end{align}
Due to the complexity of Eq.~\eqref{eq:MFPT_tot}, more insight can be obtained by focusing on some asymptotic behavior of $t_1(0)$. 
In the case of $F_{ext}=0$ and expressing the local force in the term of the free energy barrier $\Delta{\cal F}\lambda/2=-\Delta \mathcal{A}$
we obtain:
\begin{equation}
\lim_{F_{ext}\rightarrow 0}t_{1}(0)=\frac{8}{D\Delta{\cal F}^{2}}\sinh^{2}\left(\frac{\lambda\Delta{\cal F}}{4}\right)=\frac{\lambda^{2}}{D}\frac{\cosh\left(\beta\Delta \mathcal{A}\right)-1}{\left(\beta\Delta \mathcal{A}\right)^{2}}
\label{t1-smallf}
\end{equation}
that is in agreement with Ref.\cite{Malgaretti2016}.\\
On the other hand, for large forces, $\beta F_{ext}\lambda\gg1$, $t_1(0)$ reduces to 
\begin{equation}
t_{1}(0)=\frac{\lambda^2}{D(\beta F_{ext}\lambda)}
\label{t1-largef}
\end{equation}
Alternatively, for flat channels, $\Delta{\cal F}=0$, the MFPT  reads:
\begin{equation}
t_{1}(0)=\frac{\lambda^{2}}{D(\beta F_{ext}\lambda)^{2}}\left[e^{-\beta F_{ext}\lambda}-1+\beta F_{ext}\lambda\right]
\end{equation}

\subsection{List of symbols and abbreviations}
\begin{tabular}{@{}|c|l|}
$a$ & radius of the particles\\
$A(y)$ & potential of mean force \\
$\beta$ & inverse thermal energy\\
$D(y)$ & position dependent diffusion coefficient\\
$D_0$ & position independent diffusion coefficient\\
$F_{ext}$ & external force acting on the tracer particle\\
& see Refs.~\cite{Marconi2015,Puertas2018} and Eq.\eqref{eq:DA}\\
$\gamma_{eff}$ & effective friction coefficient (see Eqs.\eqref{eq:gamma_0},\eqref{eq:gamma})\\
$\Delta\mathcal{F}$ & effective force, see Eq.\eqref{eq:DF}\\
$J$ & tracer particle flux\\
$h_{min,max}$ & minimum, maximum channel amplitude\\
$\lambda$ & channel period\\
$L_{x,z}$ & channel thickness\\
$\rho_y(y)$ & tracer particle probability distribution along $y$\\
$\rho_{L,R}$ & probability of finding the tracer particle\\
& on the left (L) or roght (R) half of the channel\\
$P(\tau)$ & probability distribution of first passage times \\
& (see Eq.\eqref{eq:P_tau})\\ 
$\Delta S$ & see Eq.\eqref{eq:DS}\\ 
$t_1(0)$ & MFPT from $y=0$ to $y=L$ (see Eq.\eqref{eq:MFPT_tot})\\
$\tau_p$ & decay time of the first passage time distribution\\
$\langle v_y(y) \rangle$ & local longitudinal velocity (see Eq.\eqref{eq:vy}) 
\end{tabular}

\section{Results}

It is well known that interacting particles (or finite-size particles) form layers close to hard walls, due to the imbalance of the osmotic pressure~\cite{Lietor2007,Sandomirski2011,Mansoori2014,Kjellander2019}. When the wall is non-planar, this layering follows the shape of the walls, and in narrow corrugated channels it provokes a non-uniform density along the channel axis \cite{Puertas2018} (see Fig.S1). As expected, the density is symmetric with respect to the channel neck, and the maximum density is located in the neck. For the narrowest channel, the density increase in the neck is rather intense, that allows us to anticipate important effects on the transport properties.

\begin{figure}
\centering
\includegraphics[scale=0.3]{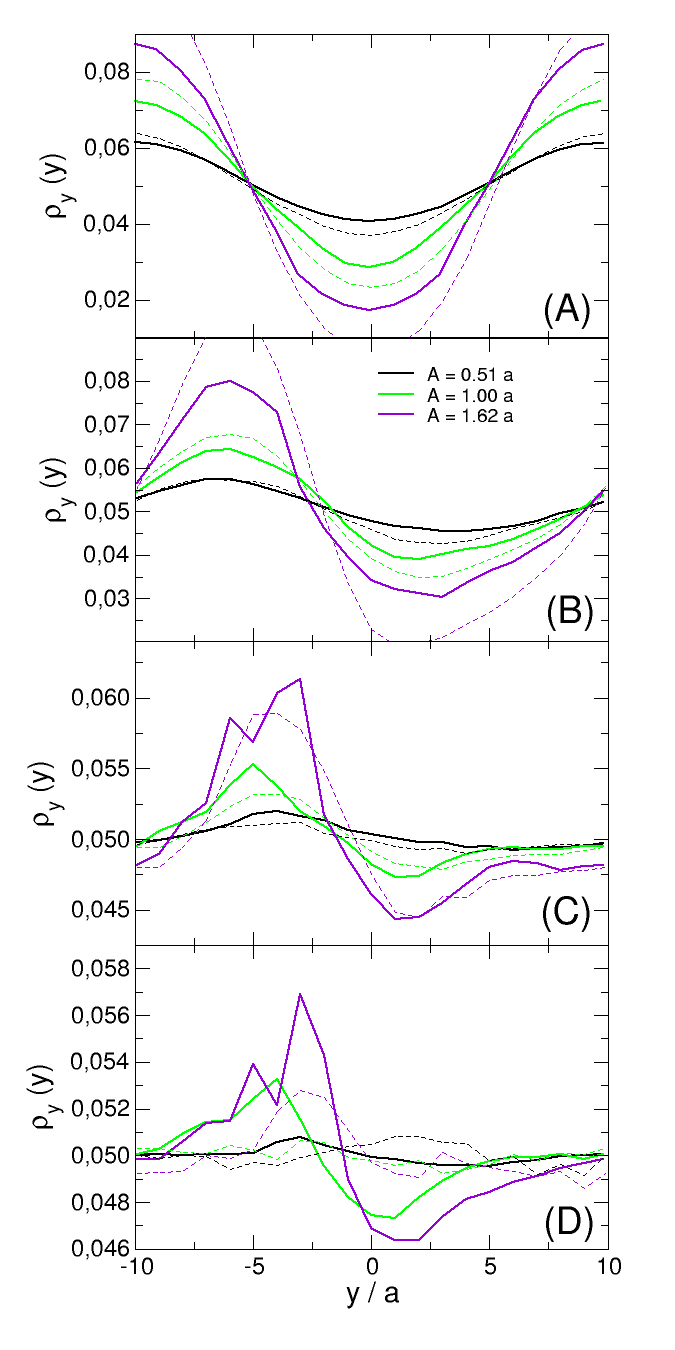}
\caption{Integrated density, $\rho_y$, as function of the  longitudinal position for different forces and corrugation amplitudes, as labeled. Solid lines represents the data for the colloidal bath, whereas dashed lines are for the thermal bath (solvent).
From top to bottom: $\beta F_{ext} \lambda=0$, $10$, $100$, $1000$. \label{rhoy-A}}
\end{figure}

The density distribution reflects the tracer position distribution at force $\beta F_{ext} \lambda=0$. When a finite external force is applied, the symmetry around the neck is broken.
In particular, such deviations from the equilibrium profile are those that eventually matter in determining the overall flux\footnote{We recall that, by definition, the equilibrium distribution leads to zero flux.}, Eq.~\eqref{eq:flux}, and hence they can be regarded as an indicator of the departure from equilibrium.
Such deviations are more quantitatively observed calculating the density integrated in slabs perpendicular to the channel axis, namely, parallel to the $XZ$-plane:

\begin{equation}
\rho_y(y)\:=\:\frac{1}{P_z}\int_{\cal V} \rho(x,y,z) dz dx
\end{equation}
\noindent where $\cal V$ is the slab volume, which we take of thickness $a$ along the longitudinal direction, and $P_z$ is a normalization constant introduced to guarantee that $\int_{-\lambda/2}^{\lambda/2} \rho_y(y) dy = 1$. The integrated density $\rho_y$ is presented in Fig.~\ref{rhoy-A} for different forces and channels. 
The most evident effect of the external force is that the symmetry with respect to the channel neck is broken: the tracer spends more time "before" the neck than "after" the neck (compare the diverse panels of Fig.~\ref{rhoy-A}). For large forces, the modulation of the density becomes less and less important (note that different scales are used in the different panels), and the tracer is more confined to the channel mid-plane. 
As shown in Fig.~\ref{rhoy-A}, the above mentioned behavior is robust and it holds for both colloidal and thermal bath (solvent). However, it is interesting to notice that, for smaller forces, the density of a tracer suspended in a thermal bath undergoes larger variations than the one for a tracer suspended in a colloidal bath whereas the opposite holds for larger forces.

\begin{figure}
\includegraphics[scale=0.35]{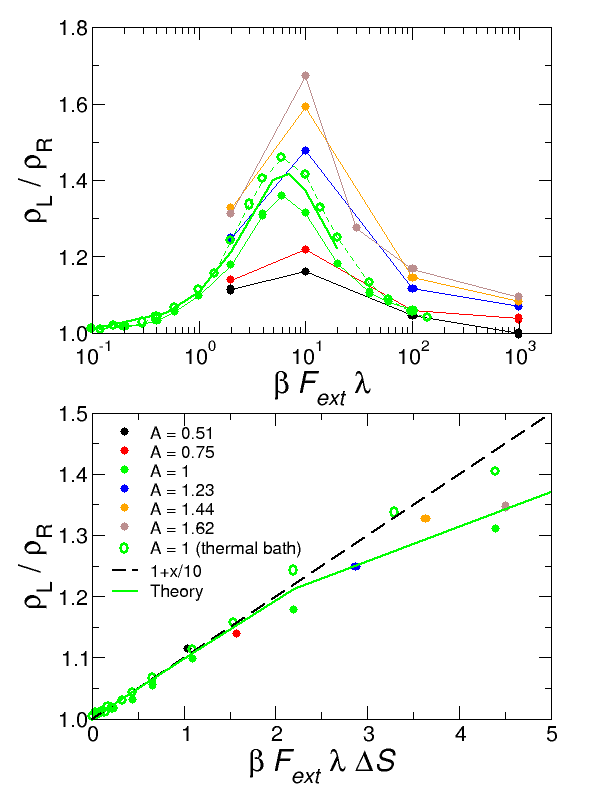}
\caption{Top: Ratio of the integrated densities in the left and right of the channel neck. The color-code is reported in the legend. The solid green line is the theoretical prediction obtained by integrating Eq.~\eqref{eq:sol-FJ} over the two halves of the channel. 
For $A=1a$, a finer grid of forces has been simulated, to locate the position of the maximum more accurately. 
Bottom: linear regime where the data collapse onto a straight line when plotted as function of $\beta F_\text{ext}\lambda \Delta S$. 
The solid green line is as in the upper panel, the solid yellow line is from Eq.~(S$29$) that assumes a piece-wise linear effective free energy,
 whereas the dashed line is the fit of Eq.~(\ref{eq:rho-ratio}) to the data.\label{fig:left-right}
}
\end{figure}

For both the thermal and colloidal baths, the effect of the external force is first to shift the modulation of the integrated density, decreasing it only slightly. Larger forces decrease it more dramatically, with small changes in the maximum and minimum positions. Whereas the former trend breaks the symmetry between both sides of the channel, the latter recovers it by eliminating the density modulation.\\
\noindent The competition of these opposing effects is captured by the ratio of the integrated densities before and after the neck, $\rho_L/\rho_R$, presented in the top panel of Fig.~\ref{fig:left-right}.
In all cases, the ratio $\rho_L/\rho_R$ displays a maximum around $\beta F_{ext} \lambda \approx 5-10$, and the ratio increases with the amplitude, $A$, of the  corrugation. 
Interestingly, the position of the maximum coincides for both ideal and colloidal baths and is also in good quantitative agreement with the prediction of the Fick-Jacobs model.

\noindent The bottom panel of Fig.~\ref{fig:left-right} shows the density ratio for small forces as a function of $\beta F_{ext} \lambda \Delta S$ compared against the theoretical predictions, green solid line (see Fig.S2 for a quantitative estimation of the mismatch between the numerical data and the theoretical predictions).
Interestingly, for small forces the data collapse on a straight line, as predicted by our simplified model, Eq.~\eqref{eq:rho-ratio} with $\mathcal{H}=1/10$.

\begin{figure}
\includegraphics[scale=0.3]{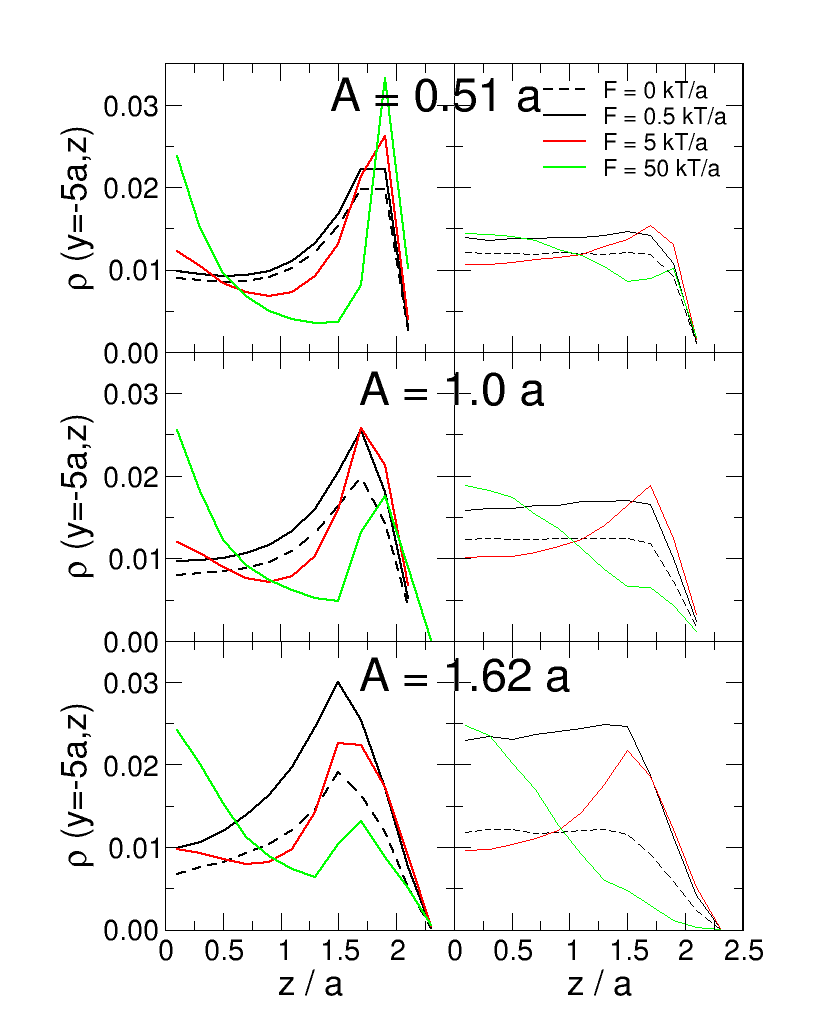}
\caption{Density in the plane $y=-5\,a$ ($\lambda/4$ before the channel
neck) for different corrugations and forces, as labeled, and the unforced tracer (dashed lines). Left panels are for the colloidal bath, right panels for the thermal bath (solvent).
\label{rho-y5}
}
\end{figure}

In order to analyze the tracer position distribution along the direction perpendicular to the force we can rely solely on numerical results since the theoretical model integrates out the transverse degrees of freedom. The variation of the tracer position distribution is studied in Fig. \ref{rho-y5}, where we focus on the plane $y=-\lambda/4=-5a$. The two extreme values of $A$, and the intermediate case are studied in the three force regimes, in addition to the unforced tracer (dashed lines). The density is normalized in all cases to give $\int_V \rho dV = 1$, where $V$ is the volume of the channel from $y=-\lambda/2$ to $y=\lambda/2$. 

As expected, in the absence of external forces the density distribution of a colloid suspended in a thermal bath (solvent) is homogeneous, whereas for a colloidal bath it shows accumulations at the walls. We remark that these accumulations are responsible for the onset of off-diagonal elements in the diffusion matrix in dense systems~\cite{Marconi2015}. In both baths, for small corrugations and forces, the distribution is only slightly modified, and it gets more affected upon increasing the value of the corrugation amplitude $A$, as anticipated by the shift of the maximum in the integrated densities in Fig.~\ref{rhoy-A}. More interestingly, the external force changes the density distribution qualitatively, increasing the tracer density at the channel mid-plane. The effect is more dramatic for the narrowest channel, where the maximum at $z=0$ surpasses the maximum at the wall. This indicates that the tracers are {\sl confined} in the center of the channel due to the neck. The tracer distribution in other planes perpendicular to the force yield qualitatively the same conclusions. 

\subsection{Tracer friction}
We study next the dynamics of the tracer, focusing first on the tracer instantaneous velocity in the direction of the force, which we integrate in slabs perpendicular to the force:

\begin{equation}
\langle v_y(y) \rangle = \dfrac{\int_{\cal V} v_y(x,y,z) \rho (x,y,z) dx dz}{\int_{\cal V} \rho(x,y,z) dx dz}
\label{eq:vy}
\end{equation}

\noindent
where $\cal V$ corresponds to the same slabs as above.  As naively expected, the velocity is lower than the average before the neck, and greater after it; the effect being more pronounced for larger channel modulations, for both baths (see Fig.(S2)). This explains the density variations along the channel, as the tracer spends more time before the neck than after it. Interestingly, while the velocity variation is symmetric for small forces, it becomes strongly asymmetric for larger ones, although the overall deviation from the average velocity decreases, corresponding to the force range where the density ratio $\rho_L/\rho_R$ decreases. Due to the collision with the bath particles, the tracer velocity is lower in the colloidal bath but the effects of the corrugation are qualitatively similar in both cases.

\begin{figure}
\includegraphics[scale=0.3]{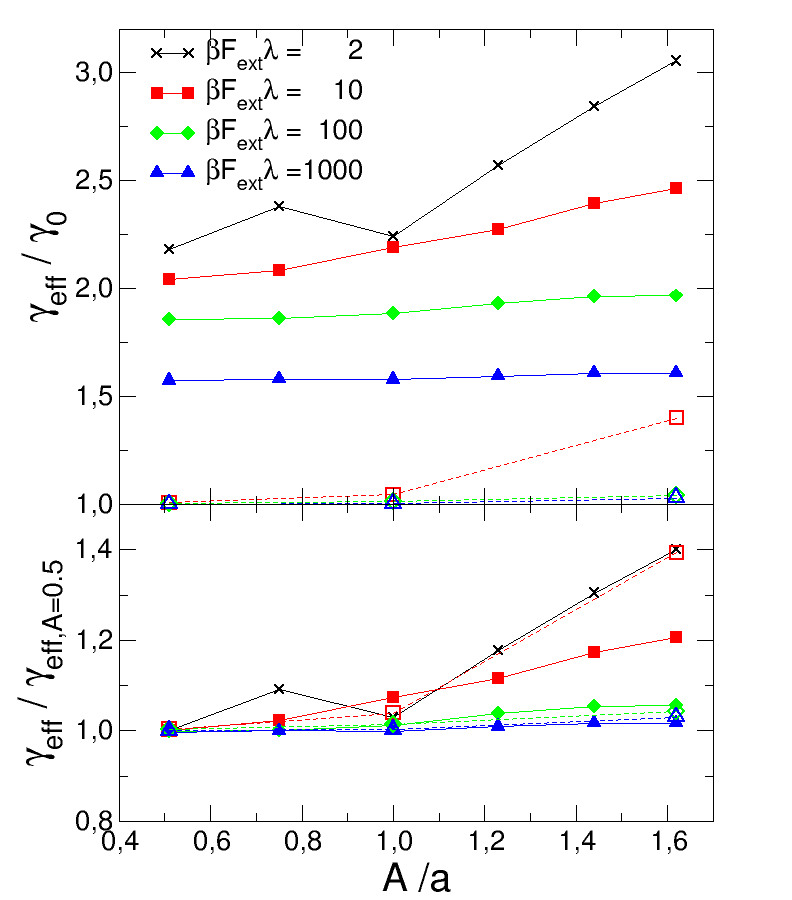}
\caption{Top: Friction coefficient, $\gamma_{eff}$, as a function of the corrugation, for different external forces, as labeled. The results for the the colloidal bath are shown with filled symbols and solid lines whereas the thermal bath (solvent) is presented with open symbols and dashed lines. 
Bottom: same as top panel but for $\gamma_{eff}$ normalized by the value attained at $A/a=0.5$.
\label{gammaA}}
\end{figure}

The tracer flux can be determined from the velocity as $J(y) = \langle v_y(y) \rangle \: \rho_y (y)$, which is independent of $y$, as expected for the stationary state. This 
determines 
the friction coefficient from the linear dependence of $J$ with $F_{ext}$
\begin{align}
\bar{\rho} F=\gammaeff J\,,
\label{eq:gamma_0}
\end{align}
where $\bar{\rho}$ is the average density. Alternatively, the friction coefficient can be calculated from the mean tracer displacement for long times, $\delta y(t)$, and defining an average tracer velocity as $\bar{v}_y = \lim_{t\rightarrow \infty}\langle \delta y(t) \rangle / t$. The latter allows us to calculate the friction coefficient using the mean field relation 
\begin{align}
F_{ext} = \gammaeff \bar{v}_y\,.
\label{eq:gamma}
\end{align}
We remark that the two definition in Eq.~\eqref{eq:gamma_0} and Eq.~\eqref{eq:gamma} coincide.
 The top panel of Fig. \ref{gammaA} shows the results of $\gammaeff$ as a function of the corrugation amplitude, for different forces and the two baths. 

In all cases, the friction coefficient decreases with the force, corresponding to the microscopic equivalent of the shear thinning observed in viscoelastic fluids (the steps in the force are probably too high to detect any non-monotonic trend in the friction coefficient, as predicted by some theoretical works\cite{Illien2014,Benichou2016,Benichou2018}). The agreement, within the statistical noise of the data, of the friction coefficient for $\beta F_{ext} \lambda=2$ and $\beta F_{ext} \lambda=10$ for small $A$ indicates that both forces are in the linear regime, while this has not been reached for large values of $A$. Interestingly, the increase of the friction coefficient at low forces is larger than at stronger pulling, resulting in more important thinning for strongly corrugated channels. This can be rationalized using the density profile perpendicular to the force (see Fig. \ref{rho-y5}), which indicates that the tracer is confined effectively to the central region of the channel for larger forces, thus passing through the neck irrespective of its width.
This phenomenology is observed also for the thermal bath, where only for large corrugations the effective friction coefficient is significantly larger than the solvent friction coefficient.
Interestingly, the lower panel of Fig.\ref{gammaA} shows that the relative increase of $\gamma_{eff}$ of colloidal and thermal bath coincide at large forces whereas for weaker pulling the thermal bath shows an enhanced sensitivity to the geometry of the channel as compared to the colloidal bath. Hence colloid-colloid interactions reduce the sensitivity of the transport coefficient to the geometry. Such a feature can be of particular interest for polydisperse colloidal suspensions where the transport coefficients of different colloids may show a diverse sensitivity to the geometry of the channel.

Finally, we study also the transversal tracer velocity, integrated also in slabs of width $a$. The general effect of the corrugated channel is to push the tracer to the midplane before the neck and out of it after the neck, making particle diffusion anisotropic \cite{Marconi2015} for both baths. This effect is enhanced for large corrugations and forces (see Fig.S3 where a negative velocity indicates "to-the-midplane", whereas a positive velocity is "out-of-the-midplane"). For small forces, the data for different corrugation amplitudes collapse onto a master curve (within the statistical noise), when the velocity is rescaled with the force, indicating a linear regime also in the transversal velocity. For larger forces, the confining before the neck is more intense for strong corrugations, as noted by the deep minimum of $v_z$ for large $A$, more intense for the thermal bath. These results are consistent with the maximum tracer density in the midplane (Fig. \ref{rho-y5}) and the concomitant independence of the friction with the corrugation amplitude (Fig. \ref{gammaA}) for large forces, and ultimately with the diffusion anisotropy observed in non-planar channels, irrespective of the presence of the colloidal bath.

\subsection{Mean First Passage times}

\begin{figure}
\includegraphics[scale=0.3]{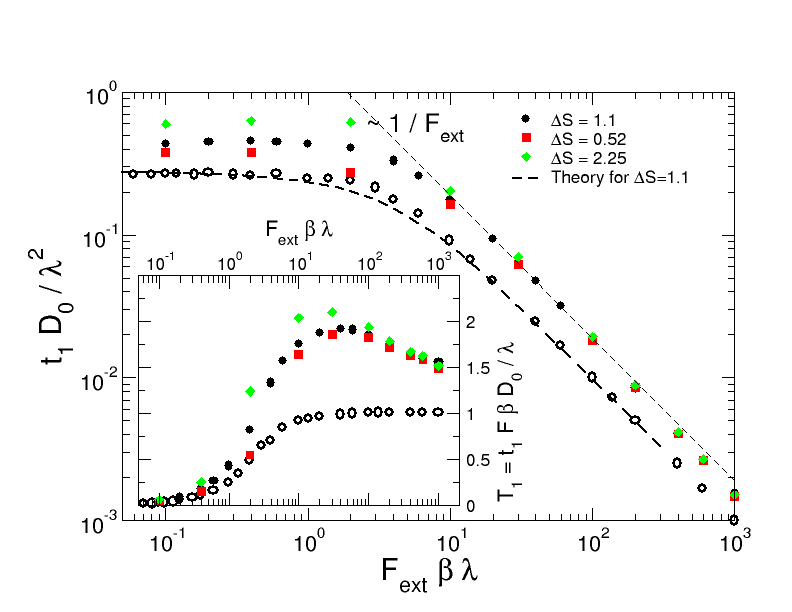}
\caption{Mean first passage times as a function of the force for different corrugations, as labeled. The open symbols show the results for the isolated tracer. \label{mfpt}}
\end{figure}

We finally study the first passage times, $\tau$, of the forced tracer. In particular we focus on the Mean First Passage Time (MFPT), $t_1=\langle \tau \rangle$, extracted from the passage time distributions (see suppl. mat.).
The MFPT is plotted in Fig. \ref{mfpt} for all forces and three corrugation amplitudes, as well as the results of the isolated tracer, which can be accurately described by the theoretical model. As expected, the presence of the colloidal bath increases the MFPT due to the increase in the effective friction, but the overall dependence of $t_1$ on the force is unaffected (compare the open and closed circles in the figure). The graph also shows that for small forces the MFPT depends also on the corrugation amplitude, while it becomes independent for large forces. This result is in agreement with the simulations of the single particle and the theoretical model, as shown by the limits for small and large force, Eqs. (\ref{t1-smallf}) and (\ref{t1-largef}). Also, the decay of the MFPT for the large forces as $\propto 1/F_{ext}$ is correctly predicted, although the prefactor is modified by the colloidal bath.

The inset of Fig.\ref{mfpt} shows the deviation of $t_1$ from the decay $\propto1/F_{ext}$ by plotting the product $t_1 F_{ext}$. Interestingly, tracers in a colloidal bath show a remarkable difference as compared to their counterparts in a purely ideal bath. Such a difference persists for a few decades and can be exploited to characterize the properties of the colloidal bath from the MFPT of the tracer.

\begin{figure}
\includegraphics[scale=0.3]{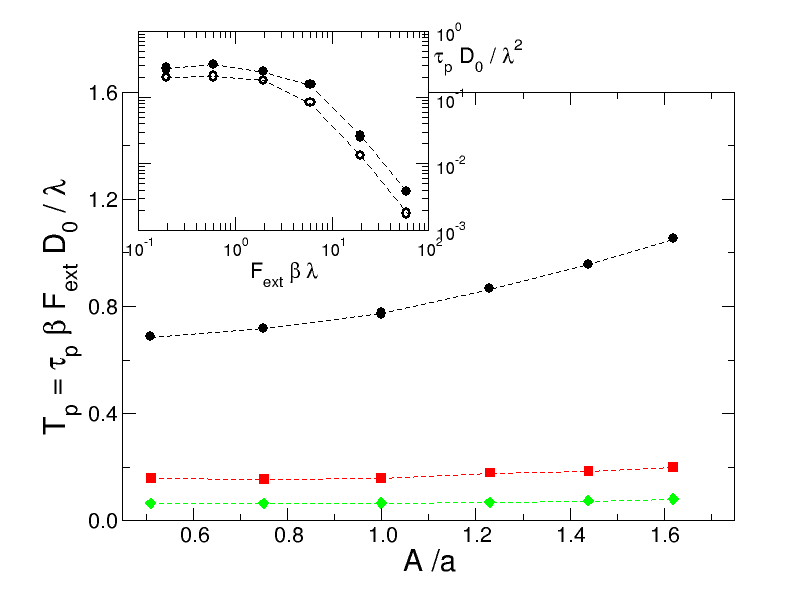}
\caption{Characteristic time as a function of the corrugation amplitude, for different forces, as labeled. The inset shows $\tau_p$ as a function of the force for the isolated tracer (open circles) and in the colloidal bath (solid circles). \label{taup}}
\end{figure}

The full distributions of first passage times $T=\tau \beta F_{ext} D_0 /\lambda$ are presented in Fig.S4 for different forces and corrugation parameters in the presence of the colloidal bath. All distributions are strongly asymmetric, with an exponential decay for large times. Comparing the passage time distributions for different forces, one notices that the decay for large times is strongly reduced for larger forces, but the maximum is almost constant when corrected with the force, as expected from our model for large forces (see eq. \ref{t1-largef}). Similar tails in the distribution of passage times are also present in the case of an isolated Brownian tracer (not shown).
A salient feature of the distribution of passage times is its  decay for large times. We have fitted this decay with an exponential form (see dashed lines in Fig.S4), 

\begin{equation}
P(\tau) \sim \exp\left\lbrace - \tau / \tau_p \right\rbrace, 
\label{eq:P_tau}
\end{equation}

\noindent to extract its characteristic time scale $\tau_p$. This is presented in Fig.~\ref{taup} for different forces and corrugations. As for the case of the MFTP, $\tau_p$ depends on the corrugation amplitude only for small forces, while it is constant for large forcings. The inset in the figure shows the evolution of $\tau_p$ with the force for a corrugation amplitude of $A=1.0 a$, comparing the tracer in the bath with an isolated tracer. Similar to the MFPT, $\tau_p$ is constant for small forces and decreases for large forces faster than $1/F$ in both cases. Again, the effect of the bath is to increase the time scales keeping the dependence on the corrugation amplitude and on the external force.

\section{Conclusions}
Microrheology has been applied to simple systems, where the models have been developed, and also to more complex systems beyond colloids, such as polymer or fiber solutions \cite{Gisler1999,Oppong2005}, soaps \cite{Prasad2009}, and particularly biological environments, such as mucus \cite{Weigand2017}, or the interior of living cells \cite{Nishizawa2017}.

In this work, we present simulations and a theoretical model of active microrheology in corrugated channels, studying the effect of the corrugation amplitude: a tracer particle is pulled with a constant force through a colloidal bath confined in the channel. The model is based on the Fick-Jacobs approximation~\cite{Jacobs1967,Zwanzig1992,Reguera2001,Kalinay2006,Kalinay2018} for an isolated tracer in a corrugated channel, and it is validated by simulations without the colloidal bath (only the solvent, as a thermal bath, is considered). The results of the simulations with the colloidal bath follow the same trends of the isolated tracer, although quantitative differences are noticed. 

In particular, the simulation results show that upon increasing the corrugation amplitude, the tracer accumulates at the channel neck: the position distribution peaks, the velocity parallel to the force decreases, and the resulting friction coefficient increases. As a result, the distribution of passage times becomes more heterogeneous for narrower bottlenecks. 
On the other hand, for strong forces, the tracer is confined effectively to the center of the channel as shown by the tracer position distribution and the transversal velocity. Thus, the friction coefficient decreases~\cite{Squires2005,Carpen2005,DePuit2011,Puertas2014,Puertas2018} and the passage time distribution narrows, with the force thinning effect being more important for large corrugations. These general trends are obtained for the colloidal bath but also for the thermal bath, indicating that the origin of this phenomenology is the confinement, and the colloidal bath modifies it only quantitatively. Strong effects are noticed, nevertheless, in the effective friction, and in particular the force thinning.

The comparison with the model is performed quantitatively by studying the ratio of the densities before and after the neck, and the mean first passage times. The former grows linearly for small forces, with a slope proportional to the corrugation parameter $\Delta S$, in agreement with the model. For larger forces, the ratio displays a maximum whose position depends only slightly on $\Delta S$. The mean first passage time shows a plateau for small forces, and decreases as the inverse of the force, also in agreement with the the model. The effect of the bath is again merely quantitative for these two parameters. 

The good agreement between the colloidal and thermal bath is the cornerstone that proves the reliability of the semi-analytical model. Indeed, the model is derived under the assumption of an ideal bath and hence it can work only when colloid-colloid interactions are disregardable. Interestingly, our work shows that the regime of reliability of the semi-analytical model extends up to moderate densities, $\phi=0.2$, of the colloidal bath. This implies that the model we propose can provide quantitatively reliable predictions about the transport of colloidal suspensions across porous media. Such a quantitatively reliable and simple model can be exploited in several biological and technological applications, spanning from mucus \cite{Weigand2017}, or the interior of living cells~\cite{Nishizawa2017} to oil recovery~\cite{Gravelle2011,Foroozesh2020,Zhang2020,Bizmarkeabc2020} and gas production  where the colloidal particle concentration can be quire low, $\phi\simeq 0.05$, see Refs.~\cite{Gravelle2011,Bizmarkeabc2020}.\\
Moreover, our findings provide a solid starting point for exploring the more complex transport regimes that may occur at higher densities~\cite{Gazuz2009,Gruber2016,Su2017}. Indeed, theoretical models have predicted the onset of non-linear response in driven dense systems~\cite{Oshanin2013}. 
In addition, we foresee possible extensions of the model, for example to account for unsteady flows or time-dependent pressure drops~\cite{Fourar2004}. \\
Finally, in many circumstances like transport in narrow channels~\cite{MalgarettiSpecial}, polymer translocation~\cite{MuthukumarBook}, or any instance that triggers an irreversible cascade of events, the stationary flux is not the quantity of interest, rather the first passage time~\cite{Rednerb} is the crucial quantity. Our comparison shows that such quantity can be reliably predicted even in the presence of particle-particle interactions and correlations, hence extending previous ones~\cite{Malgaretti2016} that did not account for interactions among particles.

\section*{Acknowledgments}
PM acknowledges funding by the Deutsche Forschungsgemeinschaft (DFG, German Research Foundation) – Project-ID 416229255 – SFB 1411. \\
AMP acknowledges financial support from the Spanish Ministerio de Ciencia, through project PGC2018-101555-B-I00 and UAL/CECEU/FEDER through project UAL18-FQM-B038-A.\\
IP acknowledges support from Ministerio de Ciencia, Innovaci\'{o}n y Universidades MCIU/AEI/FEDER for financial support under grant agreement PGC2018-098373-B-100 AEI/FEDER-EU and from Generalitat de Catalunya under project 2017SGR-884, and Swiss National Science Foundation Project No. 200021-175719.

\bibliography{biblio_PM}

\begin{thebibliography}{62}
\expandafter\ifx\csname natexlab\endcsname\relax\def\natexlab#1{#1}\fi
\expandafter\ifx\csname bibnamefont\endcsname\relax
  \def\bibnamefont#1{#1}\fi
\expandafter\ifx\csname bibfnamefont\endcsname\relax
  \def\bibfnamefont#1{#1}\fi
\expandafter\ifx\csname citenamefont\endcsname\relax
  \def\citenamefont#1{#1}\fi
\expandafter\ifx\csname url\endcsname\relax
  \def\url#1{\texttt{#1}}\fi
\expandafter\ifx\csname urlprefix\endcsname\relax\def\urlprefix{URL }\fi
\providecommand{\bibinfo}[2]{#2}
\providecommand{\eprint}[2][]{\url{#2}}

\bibitem[{\citenamefont{Furst and Squires}(2017)}]{Furst2017}
\bibinfo{author}{\bibfnamefont{E.}~\bibnamefont{Furst}} \bibnamefont{and}
  \bibinfo{author}{\bibfnamefont{T.}~\bibnamefont{Squires}},
  \emph{\bibinfo{title}{Microrheology}} (\bibinfo{publisher}{Oxford University
  Press, UK}, \bibinfo{year}{2017}).

\bibitem[{\citenamefont{Mason and Weitz}(1995)}]{Mason1995}
\bibinfo{author}{\bibfnamefont{T.~G.} \bibnamefont{Mason}} \bibnamefont{and}
  \bibinfo{author}{\bibfnamefont{D.~A.} \bibnamefont{Weitz}},
  \bibinfo{journal}{Phys. Rev. Lett.} \textbf{\bibinfo{volume}{74}},
  \bibinfo{pages}{1250} (\bibinfo{year}{1995}).

\bibitem[{\citenamefont{Cicuta and Donald}(2007)}]{Cicuta2007}
\bibinfo{author}{\bibfnamefont{P.}~\bibnamefont{Cicuta}} \bibnamefont{and}
  \bibinfo{author}{\bibfnamefont{A.}~\bibnamefont{Donald}},
  \bibinfo{journal}{Soft Matter} \textbf{\bibinfo{volume}{3}},
  \bibinfo{pages}{1449} (\bibinfo{year}{2007}).

\bibitem[{\citenamefont{Wilson and Poon}(2011)}]{Wilson2009}
\bibinfo{author}{\bibfnamefont{L.~G.} \bibnamefont{Wilson}} \bibnamefont{and}
  \bibinfo{author}{\bibfnamefont{W.~C.~K.} \bibnamefont{Poon}},
  \bibinfo{journal}{Phys. Chem. Chem. Phys.} \textbf{\bibinfo{volume}{13}},
  \bibinfo{pages}{10617} (\bibinfo{year}{2011}).

\bibitem[{\citenamefont{Habdas et~al.}(2004)\citenamefont{Habdas, Schaar,
  Levitt, and Weeks}}]{Habdas2004}
\bibinfo{author}{\bibfnamefont{P.}~\bibnamefont{Habdas}},
  \bibinfo{author}{\bibfnamefont{D.}~\bibnamefont{Schaar}},
  \bibinfo{author}{\bibfnamefont{A.~C.} \bibnamefont{Levitt}},
  \bibnamefont{and} \bibinfo{author}{\bibfnamefont{E.~R.} \bibnamefont{Weeks}},
  \bibinfo{journal}{Europhysics Letters ({EPL})} \textbf{\bibinfo{volume}{67}},
  \bibinfo{pages}{477} (\bibinfo{year}{2004}).

\bibitem[{\citenamefont{Squires and Brady}(2005)}]{Squires2005}
\bibinfo{author}{\bibfnamefont{T.}~\bibnamefont{Squires}} \bibnamefont{and}
  \bibinfo{author}{\bibfnamefont{J.~F.} \bibnamefont{Brady}},
  \bibinfo{journal}{Phys. Fluids} \textbf{\bibinfo{volume}{17}},
  \bibinfo{pages}{073101} (\bibinfo{year}{2005}).

\bibitem[{\citenamefont{Gazuz et~al.}(2009)\citenamefont{Gazuz, Puertas,
  Voigtmann, and Fuchs}}]{Gazuz2009}
\bibinfo{author}{\bibfnamefont{I.}~\bibnamefont{Gazuz}},
  \bibinfo{author}{\bibfnamefont{A.}~\bibnamefont{Puertas}},
  \bibinfo{author}{\bibfnamefont{T.}~\bibnamefont{Voigtmann}},
  \bibnamefont{and} \bibinfo{author}{\bibfnamefont{M.}~\bibnamefont{Fuchs}},
  \bibinfo{journal}{Phys. Rev. Lett.} \textbf{\bibinfo{volume}{102}},
  \bibinfo{pages}{248302} (\bibinfo{year}{2009}).

\bibitem[{\citenamefont{Puertas and Voigtmann}(2014)}]{Puertas2014}
\bibinfo{author}{\bibfnamefont{A.}~\bibnamefont{Puertas}} \bibnamefont{and}
  \bibinfo{author}{\bibfnamefont{T.}~\bibnamefont{Voigtmann}},
  \bibinfo{journal}{J. Phys.: Condens. Matter} \textbf{\bibinfo{volume}{26}},
  \bibinfo{pages}{243101} (\bibinfo{year}{2014}).

\bibitem[{\citenamefont{Zia}(2018)}]{Zia2018}
\bibinfo{author}{\bibfnamefont{R.~N.} \bibnamefont{Zia}},
  \bibinfo{journal}{Annual Review of Fluid Mechanics}
  \textbf{\bibinfo{volume}{50}}, \bibinfo{pages}{371} (\bibinfo{year}{2018}).

\bibitem[{\citenamefont{DePuit et~al.}(2011)\citenamefont{DePuit, Khair, and
  Squires}}]{DePuit2011}
\bibinfo{author}{\bibfnamefont{R.}~\bibnamefont{DePuit}},
  \bibinfo{author}{\bibfnamefont{A.}~\bibnamefont{Khair}}, \bibnamefont{and}
  \bibinfo{author}{\bibfnamefont{T.}~\bibnamefont{Squires}},
  \bibinfo{journal}{Phys. Fluids} \textbf{\bibinfo{volume}{23}},
  \bibinfo{pages}{063102} (\bibinfo{year}{2011}).

\bibitem[{\citenamefont{Chu and Zia}(2019)}]{Chu2019}
\bibinfo{author}{\bibfnamefont{H.~C.} \bibnamefont{Chu}} \bibnamefont{and}
  \bibinfo{author}{\bibfnamefont{R.~N.} \bibnamefont{Zia}},
  \bibinfo{journal}{Journal of Colloid and Interface Science}
  \textbf{\bibinfo{volume}{539}}, \bibinfo{pages}{388 } (\bibinfo{year}{2019}),
  ISSN \bibinfo{issn}{0021-9797}.

\bibitem[{\citenamefont{Gruber et~al.}(2016)\citenamefont{Gruber, Abade,
  Puertas, and Fuchs}}]{Gruber2016}
\bibinfo{author}{\bibfnamefont{M.}~\bibnamefont{Gruber}},
  \bibinfo{author}{\bibfnamefont{G.~C.} \bibnamefont{Abade}},
  \bibinfo{author}{\bibfnamefont{A.~M.} \bibnamefont{Puertas}},
  \bibnamefont{and} \bibinfo{author}{\bibfnamefont{M.}~\bibnamefont{Fuchs}},
  \bibinfo{journal}{Phys. Rev. E} \textbf{\bibinfo{volume}{94}},
  \bibinfo{pages}{042602} (\bibinfo{year}{2016}).

\bibitem[{\citenamefont{Su et~al.}(2017)\citenamefont{Su, Swan, and
  Zia}}]{Su2017}
\bibinfo{author}{\bibfnamefont{Y.}~\bibnamefont{Su}},
  \bibinfo{author}{\bibfnamefont{J.~W.} \bibnamefont{Swan}}, \bibnamefont{and}
  \bibinfo{author}{\bibfnamefont{R.~N.} \bibnamefont{Zia}},
  \bibinfo{journal}{The Journal of Chemical Physics}
  \textbf{\bibinfo{volume}{146}}, \bibinfo{pages}{124903}
  (\bibinfo{year}{2017}).

\bibitem[{\citenamefont{B\'enichou et~al.}(2014)\citenamefont{B\'enichou,
  Illien, Oshanin, Sarracino, and Voituriez}}]{Benichou2014}
\bibinfo{author}{\bibfnamefont{O.}~\bibnamefont{B\'enichou}},
  \bibinfo{author}{\bibfnamefont{P.}~\bibnamefont{Illien}},
  \bibinfo{author}{\bibfnamefont{G.}~\bibnamefont{Oshanin}},
  \bibinfo{author}{\bibfnamefont{A.}~\bibnamefont{Sarracino}},
  \bibnamefont{and}
  \bibinfo{author}{\bibfnamefont{R.}~\bibnamefont{Voituriez}},
  \bibinfo{journal}{Phys. Rev. Lett.} \textbf{\bibinfo{volume}{113}},
  \bibinfo{pages}{268002} (\bibinfo{year}{2014}).

\bibitem[{\citenamefont{Illien et~al.}(2014)\citenamefont{Illien, B\'enichou,
  Oshanin, and Voituriez}}]{Illien2014}
\bibinfo{author}{\bibfnamefont{P.}~\bibnamefont{Illien}},
  \bibinfo{author}{\bibfnamefont{O.}~\bibnamefont{B\'enichou}},
  \bibinfo{author}{\bibfnamefont{G.}~\bibnamefont{Oshanin}}, \bibnamefont{and}
  \bibinfo{author}{\bibfnamefont{R.}~\bibnamefont{Voituriez}},
  \bibinfo{journal}{Phys. Rev. Lett.} \textbf{\bibinfo{volume}{113}},
  \bibinfo{pages}{030603} (\bibinfo{year}{2014}).

\bibitem[{\citenamefont{B\'enichou et~al.}(2016)\citenamefont{B\'enichou,
  Illien, Oshanin, Sarracino, and Voituriez}}]{Benichou2016}
\bibinfo{author}{\bibfnamefont{O.}~\bibnamefont{B\'enichou}},
  \bibinfo{author}{\bibfnamefont{P.}~\bibnamefont{Illien}},
  \bibinfo{author}{\bibfnamefont{G.}~\bibnamefont{Oshanin}},
  \bibinfo{author}{\bibfnamefont{A.}~\bibnamefont{Sarracino}},
  \bibnamefont{and}
  \bibinfo{author}{\bibfnamefont{R.}~\bibnamefont{Voituriez}},
  \bibinfo{journal}{Phys. Rev. E} \textbf{\bibinfo{volume}{93}},
  \bibinfo{pages}{032128} (\bibinfo{year}{2016}).

\bibitem[{\citenamefont{B{\'{e}}nichou
  et~al.}(2018)\citenamefont{B{\'{e}}nichou, Illien, Oshanin, Sarracino, and
  Voituriez}}]{Benichou2018}
\bibinfo{author}{\bibfnamefont{O.}~\bibnamefont{B{\'{e}}nichou}},
  \bibinfo{author}{\bibfnamefont{P.}~\bibnamefont{Illien}},
  \bibinfo{author}{\bibfnamefont{G.}~\bibnamefont{Oshanin}},
  \bibinfo{author}{\bibfnamefont{A.}~\bibnamefont{Sarracino}},
  \bibnamefont{and}
  \bibinfo{author}{\bibfnamefont{R.}~\bibnamefont{Voituriez}},
  \bibinfo{journal}{Journal of Physics: Condensed Matter}
  \textbf{\bibinfo{volume}{30}}, \bibinfo{pages}{443001}
  (\bibinfo{year}{2018}),
  \urlprefix\url{https://doi.org/10.1088/1361-648x/aae13a}.

\bibitem[{\citenamefont{Gisler and Weitz}(1999)}]{Gisler1999}
\bibinfo{author}{\bibfnamefont{T.}~\bibnamefont{Gisler}} \bibnamefont{and}
  \bibinfo{author}{\bibfnamefont{D.~A.} \bibnamefont{Weitz}},
  \bibinfo{journal}{Phys. Rev. Lett.} \textbf{\bibinfo{volume}{82}},
  \bibinfo{pages}{1606} (\bibinfo{year}{1999}).

\bibitem[{\citenamefont{Oppong et~al.}(2006)\citenamefont{Oppong, Rubatat,
  Frisken, Bailey, and de~Bruyn}}]{Oppong2005}
\bibinfo{author}{\bibfnamefont{F.~K.} \bibnamefont{Oppong}},
  \bibinfo{author}{\bibfnamefont{L.}~\bibnamefont{Rubatat}},
  \bibinfo{author}{\bibfnamefont{B.~J.} \bibnamefont{Frisken}},
  \bibinfo{author}{\bibfnamefont{A.~E.} \bibnamefont{Bailey}},
  \bibnamefont{and} \bibinfo{author}{\bibfnamefont{J.~R.}
  \bibnamefont{de~Bruyn}}, \bibinfo{journal}{Phys. Rev. E}
  \textbf{\bibinfo{volume}{73}}, \bibinfo{pages}{041405}
  (\bibinfo{year}{2006}).

\bibitem[{\citenamefont{Prasad and Weeks}(2009)}]{Prasad2009}
\bibinfo{author}{\bibfnamefont{V.}~\bibnamefont{Prasad}} \bibnamefont{and}
  \bibinfo{author}{\bibfnamefont{E.~R.} \bibnamefont{Weeks}},
  \bibinfo{journal}{Phys. Rev. Lett.} \textbf{\bibinfo{volume}{102}},
  \bibinfo{pages}{178302} (\bibinfo{year}{2009}).

\bibitem[{\citenamefont{Weigand et~al.}(2017)\citenamefont{Weigand, Messmore,
  Tu, Morales-Sanz, D.L., Beheyn, J.S., and Robertson-Anderson}}]{Weigand2017}
\bibinfo{author}{\bibfnamefont{W.}~\bibnamefont{Weigand}},
  \bibinfo{author}{\bibfnamefont{A.}~\bibnamefont{Messmore}},
  \bibinfo{author}{\bibfnamefont{J.}~\bibnamefont{Tu}},
  \bibinfo{author}{\bibfnamefont{A.}~\bibnamefont{Morales-Sanz}},
  \bibinfo{author}{\bibfnamefont{B.}~\bibnamefont{D.L.}},
  \bibinfo{author}{\bibfnamefont{D.}~\bibnamefont{Beheyn}},
  \bibinfo{author}{\bibfnamefont{U.}~\bibnamefont{J.S.}}, \bibnamefont{and}
  \bibinfo{author}{\bibfnamefont{R.}~\bibnamefont{Robertson-Anderson}},
  \bibinfo{journal}{PLoS One} \textbf{\bibinfo{volume}{12}},
  \bibinfo{pages}{e0176732} (\bibinfo{year}{2017}).

\bibitem[{\citenamefont{Nishizawa et~al.}(2017)\citenamefont{Nishizawa,
  Bremerich, Ayade, Schmidt, Ariga, and Mizuno}}]{Nishizawa2017}
\bibinfo{author}{\bibfnamefont{K.}~\bibnamefont{Nishizawa}},
  \bibinfo{author}{\bibfnamefont{M.}~\bibnamefont{Bremerich}},
  \bibinfo{author}{\bibfnamefont{H.}~\bibnamefont{Ayade}},
  \bibinfo{author}{\bibfnamefont{C.~F.} \bibnamefont{Schmidt}},
  \bibinfo{author}{\bibfnamefont{T.}~\bibnamefont{Ariga}}, \bibnamefont{and}
  \bibinfo{author}{\bibfnamefont{D.}~\bibnamefont{Mizuno}},
  \bibinfo{journal}{Science Advances} \textbf{\bibinfo{volume}{3}}
  (\bibinfo{year}{2017}).

\bibitem[{\citenamefont{Marconi et~al.}(2015)\citenamefont{Marconi, Malgaretti,
  and Pagonabarraga}}]{Marconi2015}
\bibinfo{author}{\bibfnamefont{U.}~\bibnamefont{Marconi}},
  \bibinfo{author}{\bibfnamefont{P.}~\bibnamefont{Malgaretti}},
  \bibnamefont{and}
  \bibinfo{author}{\bibfnamefont{I.}~\bibnamefont{Pagonabarraga}},
  \bibinfo{journal}{J. Chem. Phys.} \textbf{\bibinfo{volume}{134}},
  \bibinfo{pages}{184501} (\bibinfo{year}{2015}).

\bibitem[{\citenamefont{Puertas et~al.}(2018)\citenamefont{Puertas, Malgaretti,
  and Pagonabarraga}}]{Puertas2018}
\bibinfo{author}{\bibfnamefont{A.~M.} \bibnamefont{Puertas}},
  \bibinfo{author}{\bibfnamefont{P.}~\bibnamefont{Malgaretti}},
  \bibnamefont{and}
  \bibinfo{author}{\bibfnamefont{I.}~\bibnamefont{Pagonabarraga}},
  \bibinfo{journal}{The Journal of Chemical Physics}
  \textbf{\bibinfo{volume}{149}}, \bibinfo{pages}{174908}
  (\bibinfo{year}{2018}).

\bibitem[{\citenamefont{Jacobs}(1967)}]{Jacobs1967}
\bibinfo{author}{\bibfnamefont{M.}~\bibnamefont{Jacobs}},
  \emph{\bibinfo{title}{Diffusion Processes}} (\bibinfo{publisher}{Springer,
  New York}, \bibinfo{year}{1967}).

\bibitem[{\citenamefont{Zwanzig}(1992)}]{Zwanzig1992}
\bibinfo{author}{\bibfnamefont{R.}~\bibnamefont{Zwanzig}}, \bibinfo{journal}{J.
  Phys. Chem.} \textbf{\bibinfo{volume}{96}}, \bibinfo{pages}{3926}
  (\bibinfo{year}{1992}).

\bibitem[{\citenamefont{D.~Reguera}(2001)}]{Reguera2001}
\bibinfo{author}{\bibfnamefont{J.~R.} \bibnamefont{D.~Reguera}},
  \bibinfo{journal}{Phys. Rev. E} \textbf{\bibinfo{volume}{64}},
  \bibinfo{pages}{061106} (\bibinfo{year}{2001}).

\bibitem[{\citenamefont{Kalinay and Percus}(2006)}]{Kalinay2006}
\bibinfo{author}{\bibfnamefont{P.}~\bibnamefont{Kalinay}} \bibnamefont{and}
  \bibinfo{author}{\bibfnamefont{J.~K.} \bibnamefont{Percus}},
  \bibinfo{journal}{Phys. Rev. E} \textbf{\bibinfo{volume}{74}},
  \bibinfo{pages}{049904} (\bibinfo{year}{2006}).

\bibitem[{\citenamefont{Pineda et~al.}(2012)\citenamefont{Pineda,
  Alvarez-Ramirez, and Dagdug}}]{Dagdug2012_2}
\bibinfo{author}{\bibfnamefont{I.}~\bibnamefont{Pineda}},
  \bibinfo{author}{\bibfnamefont{J.}~\bibnamefont{Alvarez-Ramirez}},
  \bibnamefont{and} \bibinfo{author}{\bibfnamefont{L.}~\bibnamefont{Dagdug}},
  \bibinfo{journal}{J. Chem. Phys.} \textbf{\bibinfo{volume}{137}},
  \bibinfo{pages}{174103} (\bibinfo{year}{2012}).

\bibitem[{\citenamefont{Berezhkovskii et~al.}(2015)\citenamefont{Berezhkovskii,
  Dagdug, and Bezrukov}}]{dagdug2015}
\bibinfo{author}{\bibfnamefont{A.~M.} \bibnamefont{Berezhkovskii}},
  \bibinfo{author}{\bibfnamefont{L.}~\bibnamefont{Dagdug}}, \bibnamefont{and}
  \bibinfo{author}{\bibfnamefont{S.~M.} \bibnamefont{Bezrukov}},
  \bibinfo{journal}{The Journal of Chemical Physics}
  \textbf{\bibinfo{volume}{143}}, \bibinfo{eid}{164102} (\bibinfo{year}{2015}).

\bibitem[{\citenamefont{Kalinay and Slanina}(2018)}]{Kalinay2018}
\bibinfo{author}{\bibfnamefont{P.}~\bibnamefont{Kalinay}} \bibnamefont{and}
  \bibinfo{author}{\bibfnamefont{F.}~\bibnamefont{Slanina}},
  \bibinfo{journal}{Journal of Physics: Condensed Matter}
  \textbf{\bibinfo{volume}{30}}, \bibinfo{pages}{244002}
  (\bibinfo{year}{2018}).

\bibitem[{\citenamefont{Malgaretti et~al.}(2013)\citenamefont{Malgaretti,
  Pagonabarraga, and Rubi}}]{Malgaretti2013}
\bibinfo{author}{\bibfnamefont{P.}~\bibnamefont{Malgaretti}},
  \bibinfo{author}{\bibfnamefont{I.}~\bibnamefont{Pagonabarraga}},
  \bibnamefont{and} \bibinfo{author}{\bibfnamefont{J.}~\bibnamefont{Rubi}},
  \bibinfo{journal}{Frontiers in Physics} \textbf{\bibinfo{volume}{1}},
  \bibinfo{pages}{21} (\bibinfo{year}{2013}).

\bibitem[{\citenamefont{Martens et~al.}(2011)\citenamefont{Martens, Schmid,
  Schimansky-Geier, and H\"anggi}}]{Martens2011}
\bibinfo{author}{\bibfnamefont{S.}~\bibnamefont{Martens}},
  \bibinfo{author}{\bibfnamefont{G.}~\bibnamefont{Schmid}},
  \bibinfo{author}{\bibfnamefont{L.}~\bibnamefont{Schimansky-Geier}},
  \bibnamefont{and} \bibinfo{author}{\bibfnamefont{P.}~\bibnamefont{H\"anggi}},
  \bibinfo{journal}{Phys. Rev. E} \textbf{\bibinfo{volume}{83}},
  \bibinfo{pages}{051135} (\bibinfo{year}{2011}).

\bibitem[{\citenamefont{Martens et~al.}(2013)\citenamefont{Martens, Straube,
  Schmid, Schimansky-Geier, and H\"anggi}}]{Hanggi2013}
\bibinfo{author}{\bibfnamefont{S.}~\bibnamefont{Martens}},
  \bibinfo{author}{\bibfnamefont{A.~V.} \bibnamefont{Straube}},
  \bibinfo{author}{\bibfnamefont{G.}~\bibnamefont{Schmid}},
  \bibinfo{author}{\bibfnamefont{L.}~\bibnamefont{Schimansky-Geier}},
  \bibnamefont{and} \bibinfo{author}{\bibfnamefont{P.}~\bibnamefont{H\"anggi}},
  \bibinfo{journal}{Phys. Rev. Lett.} \textbf{\bibinfo{volume}{110}},
  \bibinfo{pages}{010601} (\bibinfo{year}{2013}).

\bibitem[{\citenamefont{Malgaretti et~al.}(2014)\citenamefont{Malgaretti,
  Pagonabarraga, and Rubi}}]{Malgaretti2014}
\bibinfo{author}{\bibfnamefont{P.}~\bibnamefont{Malgaretti}},
  \bibinfo{author}{\bibfnamefont{I.}~\bibnamefont{Pagonabarraga}},
  \bibnamefont{and} \bibinfo{author}{\bibfnamefont{J.~M.} \bibnamefont{Rubi}},
  \bibinfo{journal}{Phys. Rev. Lett} \textbf{\bibinfo{volume}{113}},
  \bibinfo{pages}{128301} (\bibinfo{year}{2014}).

\bibitem[{\citenamefont{Chinappi and Malgaretti}(2018)}]{Chinappi2018}
\bibinfo{author}{\bibfnamefont{M.}~\bibnamefont{Chinappi}} \bibnamefont{and}
  \bibinfo{author}{\bibfnamefont{P.}~\bibnamefont{Malgaretti}},
  \bibinfo{journal}{Soft Matter} \textbf{\bibinfo{volume}{14}},
  \bibinfo{pages}{9083} (\bibinfo{year}{2018}).

\bibitem[{\citenamefont{Malgaretti
  et~al.}(2019{\natexlab{a}})\citenamefont{Malgaretti, Janssen, Pagonabarraga,
  and Rubi}}]{Malgaretti2019JCP}
\bibinfo{author}{\bibfnamefont{P.}~\bibnamefont{Malgaretti}},
  \bibinfo{author}{\bibfnamefont{M.}~\bibnamefont{Janssen}},
  \bibinfo{author}{\bibfnamefont{I.}~\bibnamefont{Pagonabarraga}},
  \bibnamefont{and} \bibinfo{author}{\bibfnamefont{J.~M.} \bibnamefont{Rubi}},
  \bibinfo{journal}{J. Chem. Phys.} \textbf{\bibinfo{volume}{151}},
  \bibinfo{pages}{084902} (\bibinfo{year}{2019}{\natexlab{a}}).

\bibitem[{\citenamefont{Kalinay}(2020)}]{Kalinay2020}
\bibinfo{author}{\bibfnamefont{P.}~\bibnamefont{Kalinay}},
  \bibinfo{journal}{Phys. Rev. E} \textbf{\bibinfo{volume}{102}},
  \bibinfo{pages}{042606} (\bibinfo{year}{2020}).

\bibitem[{\citenamefont{Bianco and Malgaretti}(2016)}]{Bianco2016}
\bibinfo{author}{\bibfnamefont{V.}~\bibnamefont{Bianco}} \bibnamefont{and}
  \bibinfo{author}{\bibfnamefont{P.}~\bibnamefont{Malgaretti}},
  \bibinfo{journal}{J. Chem. Phys.} \textbf{\bibinfo{volume}{145}},
  \bibinfo{pages}{114904} (\bibinfo{year}{2016}).

\bibitem[{\citenamefont{Malgaretti and Oshanin}(2019)}]{Malgaretti2019}
\bibinfo{author}{\bibfnamefont{P.}~\bibnamefont{Malgaretti}} \bibnamefont{and}
  \bibinfo{author}{\bibfnamefont{G.}~\bibnamefont{Oshanin}},
  \bibinfo{journal}{Polymers} \textbf{\bibinfo{volume}{11}}
  (\bibinfo{year}{2019}).

\bibitem[{\citenamefont{Carusela et~al.}(2021)\citenamefont{Carusela,
  Malgaretti, and Rubi}}]{Carusela2021}
\bibinfo{author}{\bibfnamefont{M.~F.} \bibnamefont{Carusela}},
  \bibinfo{author}{\bibfnamefont{P.}~\bibnamefont{Malgaretti}},
  \bibnamefont{and} \bibinfo{author}{\bibfnamefont{J.~M.} \bibnamefont{Rubi}},
  \bibinfo{journal}{Phys. Rev. E} \textbf{\bibinfo{volume}{103}},
  \bibinfo{pages}{062102} (\bibinfo{year}{2021}).

\bibitem[{\citenamefont{Malgaretti and Harting}(2021)}]{Malgaretti2021}
\bibinfo{author}{\bibfnamefont{P.}~\bibnamefont{Malgaretti}} \bibnamefont{and}
  \bibinfo{author}{\bibfnamefont{J.}~\bibnamefont{Harting}},
  \bibinfo{journal}{Soft Matter} \textbf{\bibinfo{volume}{17}},
  \bibinfo{pages}{2062} (\bibinfo{year}{2021}).

\bibitem[{\citenamefont{Ledesma-Dur{\'a}n
  et~al.}(2016)\citenamefont{Ledesma-Dur{\'a}n, Hern{\'a}ndez-Hern{\'a}ndez,
  and Santamar{\'i}a-Holek}}]{Ledesma-Duran2016}
\bibinfo{author}{\bibfnamefont{A.}~\bibnamefont{Ledesma-Dur{\'a}n}},
  \bibinfo{author}{\bibfnamefont{S.~I.}
  \bibnamefont{Hern{\'a}ndez-Hern{\'a}ndez}}, \bibnamefont{and}
  \bibinfo{author}{\bibfnamefont{I.}~\bibnamefont{Santamar{\'i}a-Holek}},
  \bibinfo{journal}{The Journal of Physical Chemistry C}
  \textbf{\bibinfo{volume}{120}}, \bibinfo{pages}{7810} (\bibinfo{year}{2016}).

\bibitem[{\citenamefont{Chac\'on-Acosta
  et~al.}(2020)\citenamefont{Chac\'on-Acosta, N\'u\~nez L\'opez, and
  Pineda}}]{Chacon-Acosta2020}
\bibinfo{author}{\bibfnamefont{G.}~\bibnamefont{Chac\'on-Acosta}},
  \bibinfo{author}{\bibfnamefont{M.}~\bibnamefont{N\'u\~nez L\'opez}},
  \bibnamefont{and} \bibinfo{author}{\bibfnamefont{I.}~\bibnamefont{Pineda}},
  \bibinfo{journal}{The Journal of Chemical Physics}
  \textbf{\bibinfo{volume}{152}}, \bibinfo{pages}{024101}
  (\bibinfo{year}{2020}).

\bibitem[{\citenamefont{Malgaretti et~al.}(2016)\citenamefont{Malgaretti,
  Pagonabarraga, and Rubi}}]{Malgaretti2016}
\bibinfo{author}{\bibfnamefont{P.}~\bibnamefont{Malgaretti}},
  \bibinfo{author}{\bibfnamefont{I.}~\bibnamefont{Pagonabarraga}},
  \bibnamefont{and} \bibinfo{author}{\bibfnamefont{J.~M.} \bibnamefont{Rubi}},
  \bibinfo{journal}{J. Chem. Phys.} \textbf{\bibinfo{volume}{114}},
  \bibinfo{pages}{034901} (\bibinfo{year}{2016}).

\bibitem[{\citenamefont{Carpen and Brady}(2005)}]{Carpen2005}
\bibinfo{author}{\bibfnamefont{I.~C.} \bibnamefont{Carpen}} \bibnamefont{and}
  \bibinfo{author}{\bibfnamefont{J.~F.} \bibnamefont{Brady}},
  \bibinfo{journal}{Journal of Rheology} \textbf{\bibinfo{volume}{49}},
  \bibinfo{pages}{1483} (\bibinfo{year}{2005}).

\bibitem[{\citenamefont{Metzler et~al.}(2014)\citenamefont{Metzler, Oshanin,
  and Redner}}]{Rednerb}
\bibinfo{author}{\bibfnamefont{R.}~\bibnamefont{Metzler}},
  \bibinfo{author}{\bibfnamefont{G.}~\bibnamefont{Oshanin}}, \bibnamefont{and}
  \bibinfo{author}{\bibfnamefont{S.}~\bibnamefont{Redner}},
  \emph{\bibinfo{title}{First-Passage Phenomena and Their Applications}}
  (\bibinfo{publisher}{World Scientific Publishers},
  \bibinfo{address}{Singapore}, \bibinfo{year}{2014}).

\bibitem[{\citenamefont{Redner}(2001)}]{Redner}
\bibinfo{author}{\bibfnamefont{S.}~\bibnamefont{Redner}},
  \emph{\bibinfo{title}{A guide to first passage processes}}
  (\bibinfo{publisher}{Cambridge University Press},
  \bibinfo{address}{Cambridge}, \bibinfo{year}{2001}).

\bibitem[{\citenamefont{Gardiner}(1994)}]{Gardiner_book}
\bibinfo{author}{\bibfnamefont{C.}~\bibnamefont{Gardiner}},
  \emph{\bibinfo{title}{Handbook of stochastic methods : for physics, chemistry
  and the natural sciences}} (\bibinfo{publisher}{Springer},
  \bibinfo{address}{Berlin}, \bibinfo{year}{1994}).

\bibitem[{\citenamefont{Li\'etor-Santos
  et~al.}(2007)\citenamefont{Li\'etor-Santos, Ch\'avez-P\'aez, M\'arquez,
  Fern\'andez-Nieves, and Medina-Noyola}}]{Lietor2007}
\bibinfo{author}{\bibfnamefont{J.~J.} \bibnamefont{Li\'etor-Santos}},
  \bibinfo{author}{\bibfnamefont{M.}~\bibnamefont{Ch\'avez-P\'aez}},
  \bibinfo{author}{\bibfnamefont{M.}~\bibnamefont{M\'arquez}},
  \bibinfo{author}{\bibfnamefont{A.}~\bibnamefont{Fern\'andez-Nieves}},
  \bibnamefont{and}
  \bibinfo{author}{\bibfnamefont{M.}~\bibnamefont{Medina-Noyola}},
  \bibinfo{journal}{Phys. Rev. E} \textbf{\bibinfo{volume}{76}},
  \bibinfo{pages}{050403} (\bibinfo{year}{2007}).

\bibitem[{\citenamefont{Sandomirski et~al.}(2011)\citenamefont{Sandomirski,
  Allahyarov, Löwen, and Egelhaaf}}]{Sandomirski2011}
\bibinfo{author}{\bibfnamefont{K.}~\bibnamefont{Sandomirski}},
  \bibinfo{author}{\bibfnamefont{E.}~\bibnamefont{Allahyarov}},
  \bibinfo{author}{\bibfnamefont{H.}~\bibnamefont{Löwen}}, \bibnamefont{and}
  \bibinfo{author}{\bibfnamefont{S.~U.} \bibnamefont{Egelhaaf}},
  \bibinfo{journal}{Soft Matter} \textbf{\bibinfo{volume}{7}},
  \bibinfo{pages}{8050} (\bibinfo{year}{2011}).

\bibitem[{\citenamefont{Mansoori and Rice}(2014)}]{Mansoori2014}
\bibinfo{author}{\bibfnamefont{G.~A.} \bibnamefont{Mansoori}} \bibnamefont{and}
  \bibinfo{author}{\bibfnamefont{S.~A.} \bibnamefont{Rice}},
  \emph{\bibinfo{title}{Confined Fluids: Structure, Properties and Phase
  Behavior}} (\bibinfo{publisher}{John Wiley and Sons, Ltd},
  \bibinfo{year}{2014}), chap.~\bibinfo{chapter}{5}, p. \bibinfo{pages}{197},
  ISBN \bibinfo{isbn}{9781118949702}.

\bibitem[{\citenamefont{Kjellander}(2019)}]{Kjellander2019}
\bibinfo{author}{\bibfnamefont{R.}~\bibnamefont{Kjellander}},
  \emph{\bibinfo{title}{Statistical Mechanics of Liquids and Solutions:
  Intermolecular Forces, Structure and Surface Interactions}}
  (\bibinfo{publisher}{CRC Press}, \bibinfo{year}{2019}), ISBN
  \bibinfo{isbn}{9781482244045}.

\bibitem[{\citenamefont{Gravelle et~al.}(2011)\citenamefont{Gravelle, Peysson,
  Tabary, and Egermann}}]{Gravelle2011}
\bibinfo{author}{\bibfnamefont{A.}~\bibnamefont{Gravelle}},
  \bibinfo{author}{\bibfnamefont{Y.}~\bibnamefont{Peysson}},
  \bibinfo{author}{\bibfnamefont{R.}~\bibnamefont{Tabary}}, \bibnamefont{and}
  \bibinfo{author}{\bibfnamefont{P.}~\bibnamefont{Egermann}},
  \bibinfo{journal}{Transport in Porous Media} \textbf{\bibinfo{volume}{88}},
  \bibinfo{pages}{441} (\bibinfo{year}{2011}).

\bibitem[{\citenamefont{Foroozesh and Kumar}(2020)}]{Foroozesh2020}
\bibinfo{author}{\bibfnamefont{J.}~\bibnamefont{Foroozesh}} \bibnamefont{and}
  \bibinfo{author}{\bibfnamefont{S.}~\bibnamefont{Kumar}},
  \bibinfo{journal}{Journal of Molecular Liquids}
  \textbf{\bibinfo{volume}{316}}, \bibinfo{pages}{113876}
  (\bibinfo{year}{2020}), ISSN \bibinfo{issn}{0167-7322}.

\bibitem[{\citenamefont{Zhang et~al.}(2020)\citenamefont{Zhang, Mohamed, Goual,
  and Piri}}]{Zhang2020}
\bibinfo{author}{\bibfnamefont{B.}~\bibnamefont{Zhang}},
  \bibinfo{author}{\bibfnamefont{A.~I.~A.} \bibnamefont{Mohamed}},
  \bibinfo{author}{\bibfnamefont{L.}~\bibnamefont{Goual}}, \bibnamefont{and}
  \bibinfo{author}{\bibfnamefont{M.}~\bibnamefont{Piri}},
  \bibinfo{journal}{Scientific Reports} \textbf{\bibinfo{volume}{10}},
  \bibinfo{pages}{17539} (\bibinfo{year}{2020}).

\bibitem[{\citenamefont{Bizmark et~al.}(2020)\citenamefont{Bizmark, Schneider,
  Priestley, and Datta}}]{Bizmarkeabc2020}
\bibinfo{author}{\bibfnamefont{N.}~\bibnamefont{Bizmark}},
  \bibinfo{author}{\bibfnamefont{J.}~\bibnamefont{Schneider}},
  \bibinfo{author}{\bibfnamefont{R.~D.} \bibnamefont{Priestley}},
  \bibnamefont{and} \bibinfo{author}{\bibfnamefont{S.~S.} \bibnamefont{Datta}},
  \bibinfo{journal}{Science Advances} \textbf{\bibinfo{volume}{6}}
  (\bibinfo{year}{2020}).

\bibitem[{\citenamefont{B\'enichou et~al.}(2013)\citenamefont{B\'enichou,
  Bodrova, Chakraborty, Illien, Law, Mej\'{\i}a-Monasterio, Oshanin, and
  Voituriez}}]{Oshanin2013}
\bibinfo{author}{\bibfnamefont{O.}~\bibnamefont{B\'enichou}},
  \bibinfo{author}{\bibfnamefont{A.}~\bibnamefont{Bodrova}},
  \bibinfo{author}{\bibfnamefont{D.}~\bibnamefont{Chakraborty}},
  \bibinfo{author}{\bibfnamefont{P.}~\bibnamefont{Illien}},
  \bibinfo{author}{\bibfnamefont{A.}~\bibnamefont{Law}},
  \bibinfo{author}{\bibfnamefont{C.}~\bibnamefont{Mej\'{\i}a-Monasterio}},
  \bibinfo{author}{\bibfnamefont{G.}~\bibnamefont{Oshanin}}, \bibnamefont{and}
  \bibinfo{author}{\bibfnamefont{R.}~\bibnamefont{Voituriez}},
  \bibinfo{journal}{Phys. Rev. Lett.} \textbf{\bibinfo{volume}{111}},
  \bibinfo{pages}{260601} (\bibinfo{year}{2013}).

\bibitem[{\citenamefont{Fourar et~al.}(2004)\citenamefont{Fourar, Radilla,
  Lenormand, and Moyne}}]{Fourar2004}
\bibinfo{author}{\bibfnamefont{M.}~\bibnamefont{Fourar}},
  \bibinfo{author}{\bibfnamefont{G.}~\bibnamefont{Radilla}},
  \bibinfo{author}{\bibfnamefont{R.}~\bibnamefont{Lenormand}},
  \bibnamefont{and} \bibinfo{author}{\bibfnamefont{C.}~\bibnamefont{Moyne}},
  \bibinfo{journal}{Advances in Water Resources} \textbf{\bibinfo{volume}{27}},
  \bibinfo{pages}{669} (\bibinfo{year}{2004}), ISSN \bibinfo{issn}{0309-1708}.

\bibitem[{\citenamefont{Malgaretti
  et~al.}(2019{\natexlab{b}})\citenamefont{Malgaretti, Oshanin, and
  Talbot}}]{MalgarettiSpecial}
\bibinfo{author}{\bibfnamefont{P.}~\bibnamefont{Malgaretti}},
  \bibinfo{author}{\bibfnamefont{G.}~\bibnamefont{Oshanin}}, \bibnamefont{and}
  \bibinfo{author}{\bibfnamefont{J.}~\bibnamefont{Talbot}},
  \bibinfo{journal}{Journal of Physics: Condensed Matter}
  \textbf{\bibinfo{volume}{31}}, \bibinfo{pages}{270201}
  (\bibinfo{year}{2019}{\natexlab{b}}).

\bibitem[{\citenamefont{Muthukumar}(2011)}]{MuthukumarBook}
\bibinfo{author}{\bibfnamefont{M.}~\bibnamefont{Muthukumar}},
  \emph{\bibinfo{title}{Polymer translocation}} (\bibinfo{publisher}{CRC
  Press}, \bibinfo{address}{Boca Raton}, \bibinfo{year}{2011}).

\bibitem[{\citenamefont{Ascher}(1998)}]{num_diff_eq}
\bibinfo{author}{\bibfnamefont{L.~R.} \bibnamefont{Ascher}, \bibfnamefont{Uri
  M. annd~Petzold}}, \emph{\bibinfo{title}{Computer Methods for Ordinary
  Differential Equations and Differential-Algebraic Equations}}
  (\bibinfo{publisher}{Society for Industrial and Applied Mathematics},
  \bibinfo{address}{Philadelphia}, \bibinfo{year}{1998}).

\end{thebibliography}

\appendix

\begin{widetext}
\section*{Simulations details}
The equation of motion for all particles is:

\begin{equation}
m \ddot{\vec{r}}_j = 
  -\gamma_0 \dot{\vec{r}}_j + \vec{\eta}_j(t) + \sum_{i\neq j} \vec{F}_{ij}
  +\vec{F}_{wall}+ \delta_{j1} \vec{F}_{ext}, \label{Langevin}
\end{equation}

\noindent where the forces acting on particle $j$ are the friction with the solvent, $\gamma_0 \dot{\vec{r}}_j$, proportional to the friction coefficient $\gamma_0$ and the particle velocity $\dot{\vec{r}}_j$, the random force $\vec{\eta}_j(t)$, which verifies the fluctuation-dissipation theorem, and the interaction forces with other particles ($\vec{F}_{ij}$) and the walls ($\vec{F}_{wall}$). An additional external force, $\vec{F}_{ext}$, pulls the tracer, labelled $j=1$; this force is constant, and acts in the $y$-direction, $\vec{F}_{ext}=F_{ext} \vec{e}_y$.

The interparticle forces between particles and with the walls derive from the continuous potentials:

\begin{equation}
V(r)=k_{\text{B}}T \left(\frac{r}{2a}\right)^{-36} \hspace{0.5cm} \mbox{and} \hspace{0.5cm}
V_w(d)=k_{\text{B}}T \left(\frac{d}{a}\right)^{-36} 
\end{equation}

\noindent respectively, where $r$ is the center to center distance between particles and $d$ is the minimum distance from the particle center to the wall. 

The confining walls are defined by the corrugated surfaces:

\begin{equation}
z=\pm \left(\frac{L_z}{2} - A \cos \frac{2\pi y}{\lambda} \right) 
\end{equation}

\noindent where $A$ and $\lambda$ are the corrugation amplitude and wavelength, respectively, and $L_z$ sets the mean separation of the walls, which we set equal to $3$ particle diameters, i.e. $L_z=6a$. 

In this work we have considered channels with different amplitudes, as shown in Table \ref{table-DeltaS}. The entropy barrier is also given, which for finite-size particles is defined as \cite{Malgaretti2016}:

\begin{equation}
\Delta S = \ln \left[\frac{h_{max}-a}{h_{min}-a}\right]
\end{equation}

\begin{table}
\centering
\begin{tabular}{c|c}
$A/a$ & $\hspace{0.5cm}\Delta S\hspace{0.5cm}$ \\ \hline
\hspace{0.5cm} 0.51 \hspace{0.5cm} & 0.522 \\
0.75 & 0.788 \\
1.0  & 1.099 \\
1.23 & 1.434 \\
1.44 & 1.815 \\
1.62 & 2.254 
\end{tabular}
\caption{Values of the corrugation amplitude, $A$, and entropy barrier, $\Delta S$, used in this work. \label{table-DeltaS}}
\end{table}

\noindent with $h_{max}$ and $h_{min}$ the maximum and minimum channel widths, respectively. In all cases, the corrugation wavelength is $\lambda=20a$, and the volume fraction of the colloidal bath is $\phi=0.20$. The simulation box has dimensions $L_x\times L_y \times L_z$, with $L_x=L_z=6a$ and $L_y=2\lambda$, resulting in $69$ particles, including the tracer. The system has periodic boundary conditions in the $XY$ and $XZ$ planes. The snapshots in Fig.1 of the main text show the systems with the lowest and highest $\Delta S$ (upper and lower panels, respectively). 

The system is equilibrated without external force and at time $t=0$, the external force is switched on. The tracer trajectory, as well as the bath average properties, are recorded. The first passage time is estimated from the time of two subsequent passages of the tracer by the neck from left to right. 

In the simulations, the length unit is the particle radius, $a$, the particle mass, $m$, is set to unit,
and energy is measured in units of $k_{\text{B}}T$. The friction coefficient with the solvent is set to $\gamma_0=10 \sqrt{k_{\text{B}}T m}/a$, which corresponds to a Brownian time (the time required to diffuse a distance equal to $a$) $\tau_B=\gamma_0/6 k_{\text{B}}T=1.67 a\sqrt{m/k_{\text{B}}T}$. The equations of motion (\ref{Langevin}) are integrated with a Heun algorithm~\cite{num_diff_eq} with a time step of $\delta t = 0.0005 a\sqrt{m/k_{\text{B}}T}$. 

\section*{Derivation of Eq.(10) of the main text}
At steady state the density profile reads:
\begin{equation}
\rho(y)=e^{-\beta A(y)}\left[-\frac{J}{D}\int_{-\frac{\lambda}{2}}^{y}e^{\beta A(z)}dz+\Pi\right]
\end{equation}
with 
\begin{align}
\beta A(y) & =\beta A_{0}(y)-\beta fy\\
\Pi & =-\frac{J}{D}\frac{\int_{-\frac{\lambda}{2}}^{\frac{\lambda}{2}}e^{\beta A(z)}dz}{e^{-\beta(A(-\lambda/2)-A(\lambda/2))}-1}=-\frac{J}{D}\Pi_{0}\\
\frac{J}{D} & =-\left[\int_{-\frac{\lambda}{2}}^{\frac{\lambda}{2}}e^{-\beta A(y)}\left[\int_{-\frac{\lambda}{2}}^{y}e^{\beta A(z)}dz+\Pi_{0}\right]dy\right]^{-1}
\end{align}
Recalling that $\beta A_{0}(y)=\beta A_{0}(y+L)$, for small forces
we have: 
\begin{eqnarray}
\Pi_{0} & \simeq & -\frac{\int_{-\frac{\lambda}{2}}^{\frac{\lambda}{2}}e^{\beta A(z)}dz}{\beta f\lambda-\frac{1}{2}\left(\beta f\lambda\right)^{2}}\\
J & \simeq & \frac{\beta Df}{\lambda}\frac{1}{Z}\\
Z & = & \frac{1}{\lambda}\int_{-\frac{\lambda}{2}}^{\frac{\lambda}{2}}e^{-\beta A_{0}(z)}dz\frac{1}{\lambda}\int_{-\frac{\lambda}{2}}^{\frac{\lambda}{2}}e^{\beta A_{0}(z)}dz
\end{eqnarray}
 Note that due to symmetry reasons we have that $J$ has no contribution
proportional to $f^{2}$. At linear order in $f$ we have: 
\begin{align}
\rho(y)\simeq&\frac{1}{Z}e^{-\beta A_{0}(y)}\left(1+\beta fy\right)\frac{1}{\lambda}\left[-\beta f\int_{-\frac{\lambda}{2}}^{y}e^{\beta A_{0}(z)}(1-\beta fz)dz\right.\nonumber\\
&\left.+\left(1+\beta f\frac{\lambda}{2}\right)\frac{1}{\lambda}\int_{-\frac{\lambda}{2}}^{\frac{\lambda}{2}}e^{\beta A_{0}(z)}(1-\beta fz)dz\right]
\end{align}
by collecting terms we obtain: 
\begin{align}
\rho(y)\simeq &\frac{1}{Z}e^{-\beta A_{0}(y)}\frac{1}{\lambda}\left[-\beta f\int_{-\frac{\lambda}{2}}^{y}e^{\beta A_{0}(z)}dz\right.\nonumber\\
&\left.+\frac{1}{\lambda}(1+\beta fy+\beta f\frac{\lambda}{2})\int_{-\frac{\lambda}{2}}^{\frac{\lambda}{2}}e^{\beta A_{0}(z)}dz-\frac{1}{\lambda}\int_{-\frac{\lambda}{2}}^{\frac{\lambda}{2}}\beta fze^{\beta A_{0}(z)}dz\right]
\end{align}
Moreover, assuming that 
\begin{equation}
\beta A_{0}(y)=\begin{cases}
-2\Delta S\frac{y}{\lambda} & y<0\\
2\Delta S\frac{y}{\lambda} & y>0
\end{cases}\label{eq:DeltaS}
\end{equation}
Hence we get: 
\begin{align}
\int_{-\frac{\lambda}{2}}^{\frac{\lambda}{2}}e^{\beta A_{0}(z)}dz & =\frac{\lambda}{\Delta S}\left[e^{\Delta S}-1\right]\\
\int_{-\frac{\lambda}{2}}^{\frac{\lambda}{2}}\beta fze^{\beta A_{0}(z)}dz & =0\\
\int_{-\frac{\lambda}{2}}^{y}e^{\beta A_{0}(z)}dz & =\Gamma(y)=\begin{cases}
\frac{\lambda}{2\Delta S}\left(e^{\Delta S}-e^{-2\Delta S\frac{y}{\lambda}}\right) & y<0\\
\frac{\lambda}{2\Delta S}\left(e^{\Delta S}-2+e^{2\Delta S\frac{y}{\lambda}}\right) & y>0
\end{cases}\\
Z & =\frac{1}{\Delta S}\left[e^{\Delta S}-1\right]\frac{1}{\Delta S}\left[1-e^{-\Delta S}\right]=\frac{2}{\Delta S^{2}}\left(\cosh\Delta S-1\right)
\end{align}
that leads 
\begin{equation}
\rho(y)\simeq\frac{1}{Z}\begin{cases}
e^{2\Delta S\frac{y}{\lambda}}\frac{1}{\lambda}\left[-\beta f\frac{\lambda}{2\Delta S}\left(e^{\Delta S}-e^{-2\Delta S\frac{y}{\lambda}}\right)+\frac{1+\beta fy+\beta f\frac{\lambda}{2}}{\Delta S}\left(e^{\Delta S}-1\right)\right] & y<0\\
e^{-2\Delta S\frac{y}{\lambda}}\frac{1}{\lambda}\left[-\beta f\frac{\lambda}{2\Delta S}\left(e^{\Delta S}-2+e^{2\Delta S\frac{y}{\lambda}}\right)+\frac{1+\beta fy+\beta f\frac{\lambda}{2}}{\Delta S}\left(e^{\Delta S}-1\right)\right] & y>0
\end{cases} \label{eq:rho-ratio-full}
\end{equation}
Expanding for small $\Delta S$ we get:
\begin{align}
\Gamma(y) & \simeq\begin{cases}
\Gamma_{-}(y)=\frac{1}{2}+\frac{y}{\lambda}+\Delta S\left[\frac{1}{4}-\left(\frac{y}{\lambda}\right)^{2}\right]+\Delta S^{2}\left[\frac{1}{12}+\frac{2}{3}\left(\frac{y}{\lambda}\right)^{3}\right]  y<0\\
\Gamma_{+}(y)=\frac{1}{2}+\frac{y}{\lambda}+\Delta S\left[\frac{1}{4}+\left(\frac{y}{\lambda}\right)^{2}\right]+\Delta S^{2}\left[\frac{1}{12}+\frac{2}{3}\left(\frac{y}{\lambda}\right)^{3}\right]  y>0
\end{cases}\\
\frac{1+\beta fy}{\Delta S}\left(e^{\Delta S}-1\right) & \simeq\left(1+\beta fy\right)\left(1+\frac{\Delta S}{2}+\frac{\Delta S^{2}}{6}\right)\\
\frac{1}{Z} & \simeq1-\frac{1}{12}\Delta S^{2}
\end{align}
and the we get:
\begin{equation}
\rho(y)\simeq\begin{cases}
\frac{1}{\lambda}\left(1+2\Delta S\frac{y}{\lambda}+2\left(\Delta S\frac{y}{\lambda}\right)^{2}\right)\cdot\\
\cdot\left[-\beta f\lambda\Gamma_{-}(y)+\left(1+\beta fy+\beta f\frac{\lambda}{2}\right)\left(1+\frac{\Delta S}{2}+\frac{\Delta S^{2}}{6}\right)\right]\left(1-\frac{\Delta S^{2}}{12}\right) y<0\\
\frac{1}{\lambda}\left(1-2\Delta S\frac{y}{\lambda}+2\left(\Delta S\frac{y}{\lambda}\right)^{2}\right)\cdot\\
\cdot\left[-\beta f\lambda\Gamma_{+}(y)+\left(1+\beta fy+\beta f\frac{\lambda}{2}\right)\left(1+\frac{\Delta S}{2}+\frac{\Delta S^{2}}{6}\right)\right]\left(1-\frac{\Delta S^{2}}{12}\right) y>0
\end{cases}
\end{equation}
that can be simplified to
\begin{equation}
\rho(y)\simeq\frac{1}{\lambda}\begin{cases}
1+\left[\frac{1}{2}+2\frac{y}{\lambda}+\beta f\lambda\left(\frac{1}{2}\frac{y}{\lambda}+\left(\frac{y}{\lambda}\right)^{2}\right)\right]\Delta S+\\
+\frac{1}{12}\left[1+12\frac{y}{\lambda}+24\left(\frac{y}{\lambda}\right)^{2}+\beta f\lambda\left(2\frac{y}{\lambda}+12\left(\frac{y}{\lambda}\right)^{2}+16\left(\frac{y}{\lambda}\right)^{3}\right)\right]\Delta S^{2} & y<0\\
1+\left[\frac{1}{2}-2\frac{y}{\lambda}+\beta f\lambda\left(\frac{1}{2}\frac{y}{\lambda}-\left(\frac{y}{\lambda}\right)^{2}\right)\right]\Delta S+\\
+\frac{1}{12}\left[1-12\frac{y}{\lambda}+24\left(\frac{y}{\lambda}\right)^{2}+\beta f\lambda\left(2\frac{y}{\lambda}-12\left(\frac{y}{\lambda}\right)^{2}+16\left(\frac{y}{\lambda}\right)^{3}\right)\right]\Delta S^{2} & y>0
\end{cases}
\end{equation}
Defining 
\begin{eqnarray}
\rho_{\lambda} & = & \int_{-\frac{\lambda}{2}}^{0}\rho(y)dy\\
\rho_{R} & = & \int_{0}^{\frac{\lambda}{2}}\rho(y)dy
\end{eqnarray}
we have:
\begin{eqnarray}
\rho_{\lambda} & = & \frac{1}{2}-\frac{1}{48}\beta f\lambda\Delta S+\mathcal{O}(\Delta S^{3})\\
\rho_{R} & = & \frac{1}{2}+\frac{1}{48}\beta f\lambda\Delta S+\mathcal{O}(\Delta S^{3})
\end{eqnarray}
for small values of $f$ we can expand the ratio: 
\begin{equation}
\frac{\rho_{\lambda}}{\rho_{R}}=1-\frac{1}{12}\beta f\lambda\Delta S
\label{eq:theo_slope}
\end{equation}
We remark that the $1/12$ prefactor is due to the linear approximation of the free energy, Eq.~\ref{eq:DeltaS}. For other free energy profile with the same free-energy barrier $\Delta S$ we expect to have:
\begin{equation}
\frac{\rho_{\lambda}}{\rho_{R}}=1-\mathcal{H}\beta f\lambda\Delta S
\end{equation}

\section*{Derivation of Eq.11 of the main text}

Starting from Eq.~(3) of the main text, it is possible to derive an equation for the MFPT~\cite{Gardiner_book}, $t(y)$, for a particle over a period:
\begin{align}
-\beta D\left(\partial_y {\cal A}(y)\right)\partial_{y}t(y)+D\partial_{y}^{2}t(y) & =-1\label{eq:MFPT}\\
\partial_{y}t(y)|_{-\lambda/2} & =0\label{eq:BC-0}\\
t(y)|_{\lambda/2} & =0\label{eq:BC-L}
\end{align}
In order to gain analytical insight in solving Eq.~\eqref{eq:MFPT}, 
it is more convenient~\cite{Malgaretti2016} to approximate the effective potential as piece-wise linear:
\begin{equation}
{\cal A}(y) =  \left\{ \begin{array}{lll}
-F_{ext}y + \Delta \mathcal{A}\frac{y}{\lambda/2} & \hspace{1cm}  -\frac{\lambda}{2} <   y < 0 \\ 
-F_{ext}y - \Delta \mathcal{A}\frac{y}{\lambda/2} & \hspace{1.32cm}  0  \leq  y < \frac{\lambda}{2}
\end{array} \right.
\end{equation}
where $\Delta \mathcal{A}=\mathcal{A}(0)-\mathcal{A}(-\lambda/2)$. Accordingly we can define the piece-wise linear force:
\begin{equation}
{\cal F}(y) =  \left\{ \begin{array}{lll}
{\cal F}_1=F_{ext} + \Delta \mathcal{F} & \hspace{1cm} & y < 0 \\ 
{\cal F}_2=F_{ext} - \Delta \mathcal{F} & \hspace{1cm} & y > 0
\end{array} \right.
\end{equation}
with $\Delta \mathcal{F}=-\frac{\Delta \mathcal{A}}{\lambda/2}$.
First, we rewrite the MFPT equation, Eq.~\eqref{eq:MFPT} as
\begin{align}
\beta D{\cal F}(y)\partial_{y}t(y)+D\partial_{y}^{2}t(y) & =-1
\end{align}
Using the auxiliary variable $\partial_{y}t(y)=T(y)$ the last equations read:
\begin{align}
\beta D{\cal F}(y)T(y)+D\partial_{y}T(y) & =-1\label{eq:T}\\
T(y)|_{-\lambda/2} & =0
\end{align}
Then we split Eq.~\eqref{eq:T} into two equations:
\begin{equation}
\begin{cases}
\beta D{\cal F}_{1}T_{1}(y)+D\partial_{y}T_{1}(y)=-1 & x<0\\
\beta D{\cal F}_{2}T_{2}(y)+D\partial_{y}T_{2}(y)=-1 & x>0
\end{cases}
\end{equation}
whose solutions read:
\begin{eqnarray}
T_{1}(y) & = & e^{-\beta {\cal F}_{1}y}\left[-\frac{1}{D}\int_{-\frac{\lambda}{2}}^{y}e^{\beta {\cal F}_{1}y'}dy'+\Pi_{1}\right]\\
T_{2}(y) & = & e^{-\beta {\cal F}_{2}y}\left[-\frac{1}{D}\int_{0}^{y}e^{\beta {\cal F}_{2}y'}dy'+\Pi_{2}\right]
\end{eqnarray}
and finally 
\begin{eqnarray}
t_{1}(y) & =  \int_{-\frac{\lambda}{2}}^{y}T_{1}(y')dy'+\Lambda_{1},\, &\text{for}\,x<0\\
t_{2}(y) & =  \int_{0}^{y}T_{2}(y')dy'+\Lambda_{2},\,&\text {for}\,x>0
\end{eqnarray}
with the continuity conditions: 
\begin{eqnarray}
t_{1}(0) & = & t_{2}(0)\\
T_{1}(0) & = & T_{2}(0)
\end{eqnarray}
Due to the B.C. (Eqs.\ref{eq:BC-0}, \ref{eq:BC-L}) we have 
\begin{align}
\Pi_{1} & =0\\
\Lambda_{2} & =-\int_{0}^{\frac{\lambda}{2}}T_{2}(x)dx
\end{align}
Imposing the continuity leads to 
\begin{eqnarray*}
\Pi_{2} & = & T_{1}(0)\\
\Lambda_{1} & = & -\left[\int_{-\frac{\lambda}{2}}^{0}T_{1}(y)dy+\int_{0}^{\frac{\lambda}{2}}T_{2}(y)dy\right]
\end{eqnarray*}
The quantity we are interested in is $t_1(0)$:
 \begin{align}
t_{1}(0) =&-\frac{\lambda^{2}}{D\left[(\beta F_\text{ext}\lambda)^{2}-(\beta\Delta{\cal F}\lambda)^{2}\right]^{2}}\left[e^{-\beta F_\text{ext}\lambda}((\beta\Delta{\cal F}\lambda)^{2}-(\beta F_\text{ext}\lambda)^{2})\right.\nonumber\\
&+\left(\beta F_\text{ext}\lambda\right)^{2}-\left(\beta F_\text{ext}\lambda\right)^{3}+3\left(\beta\Delta{\cal F}\lambda\right)^{2}+\beta F_\text{ext}\lambda\left(\beta\Delta{\cal F}\lambda\right)^{2}\nonumber\\
 & \left.+e^{-\frac{1}{2}\beta \lambda(F_\text{ext}+\Delta{\cal F})}\left(2\beta(F_\text{ext}\lambda-\Delta{\cal F}\lambda)\Delta{\cal F}\lambda-2e^{\beta \lambda\Delta{\cal F}}\Delta{\cal F}\lambda(F_\text{ext}\lambda+\Delta{\cal F}\lambda)\right)\right]
\end{align}
\clearpage
\newpage

\section*{Density of colloidal particles}
Particle density in our system of quasi-hard spheres at volume fraction $\phi=0.20$, without external forces, for the two extreme corrugation amplitudes.
\begin{figure}[h!]
\includegraphics[scale=0.58]{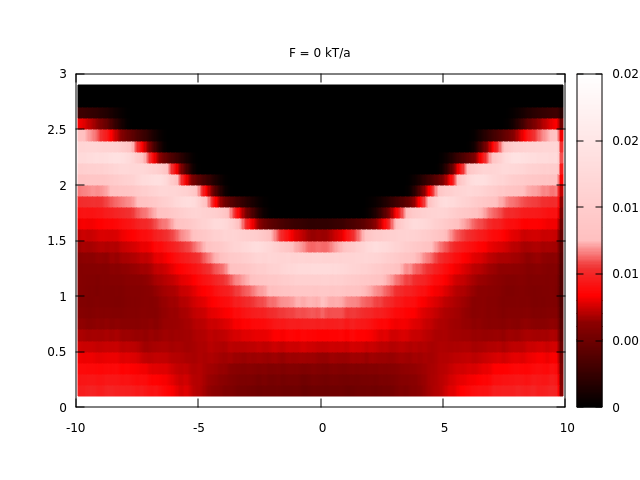}
\includegraphics[scale=0.5]{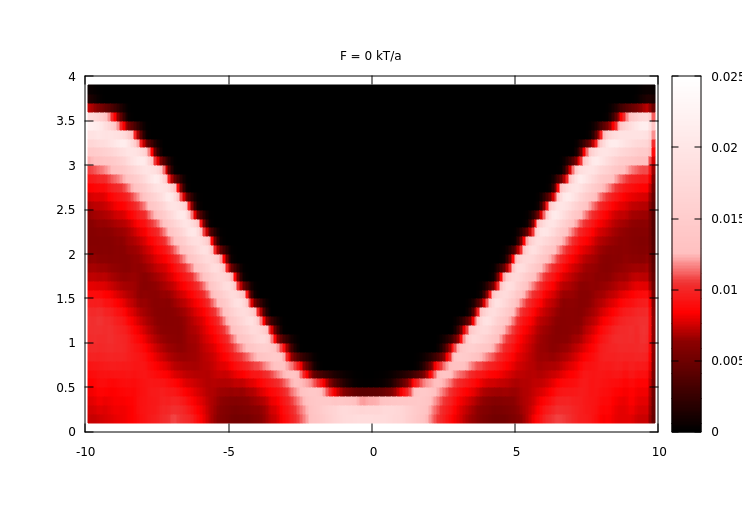}
\caption {Equilibrium ($\beta F \lambda=0$), probability density of the tracer for
$A=0.51a$ (upper panel) and $A=1.62a$ (lower panel). In both cases, $L_z=L_x=6\,a$ and $L_y=2\lambda=40\,a$.\label{rho-bath}
}
\end{figure}
\clearpage
\newpage

\section*{Theoretical/numerical comparison}
\begin{figure}[h!]
\includegraphics[scale=0.4]{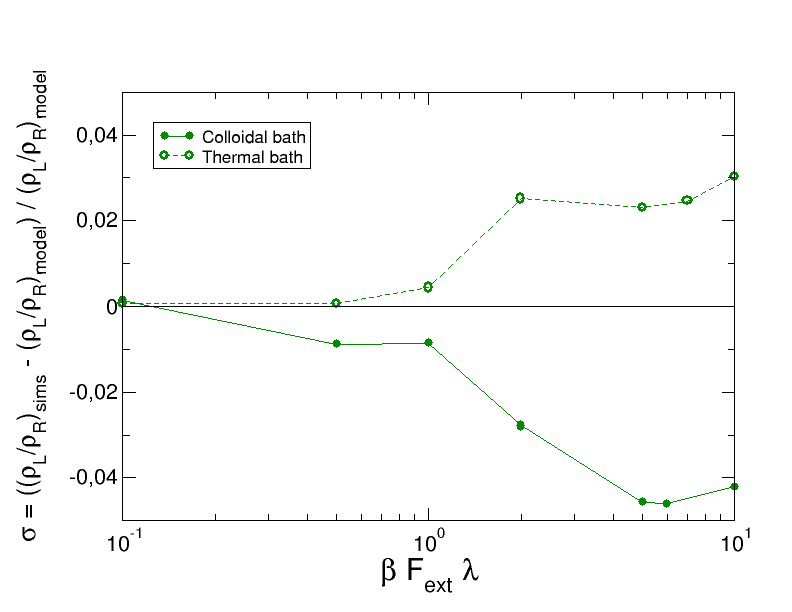}
\caption{Relative difference, $\sigma$, between the theoretical prediction of $\rho_L/\rho_R$ and the numerical one, normalized by the theoretical value, for both colloidal and thermal bath (see legend)  in the case $A/a=1$.
\label{theo-num}}
\end{figure}
\clearpage
\newpage
\section*{Longitudinal velocity of the tracer particle}
\begin{figure}[h!]
\includegraphics[scale=0.4]{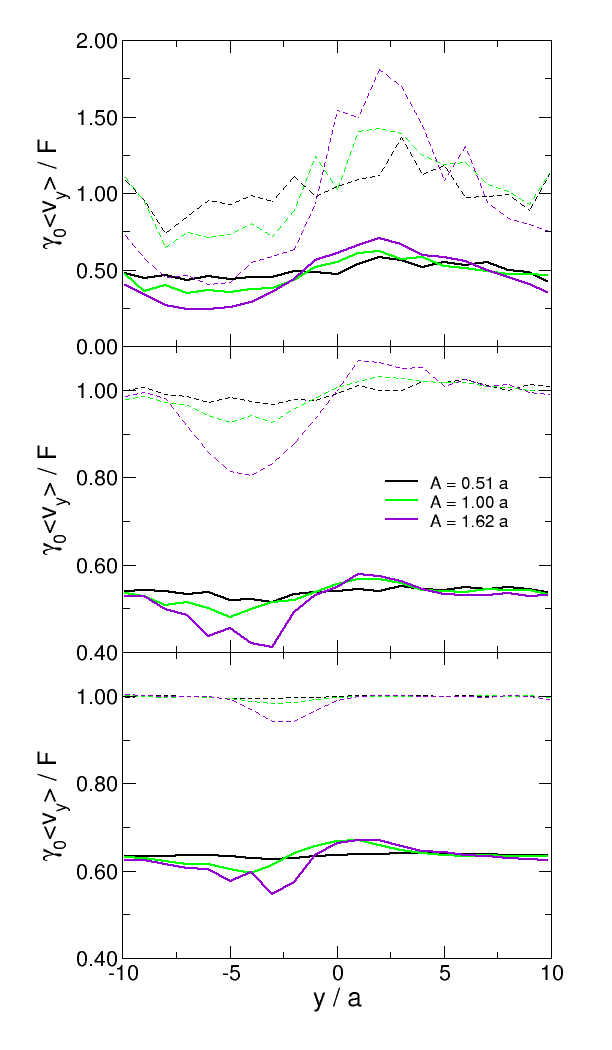}
\caption{Component of the tracer velocity along the channel. The tracer velocity is averaged in slabs transverse to the channel  direction as a function of the channel corrugation and applied external force 
for different corrugations and forces: 
$\beta F_{ext} \lambda=10$  (upper panel), 
$\beta F_{ext} \lambda=100$ (intermediate panel) and 
$\beta F_{ext}\lambda=1000$ (bottom panel). Solid lines stand for thermal bath whereas dashed lines for the colloidal bath.
\label{vy-A}}
\end{figure}
\clearpage
\newpage

\section*{Transverse velocity of the tracer particle}
\begin{figure}[h!]
\includegraphics[scale=0.4]{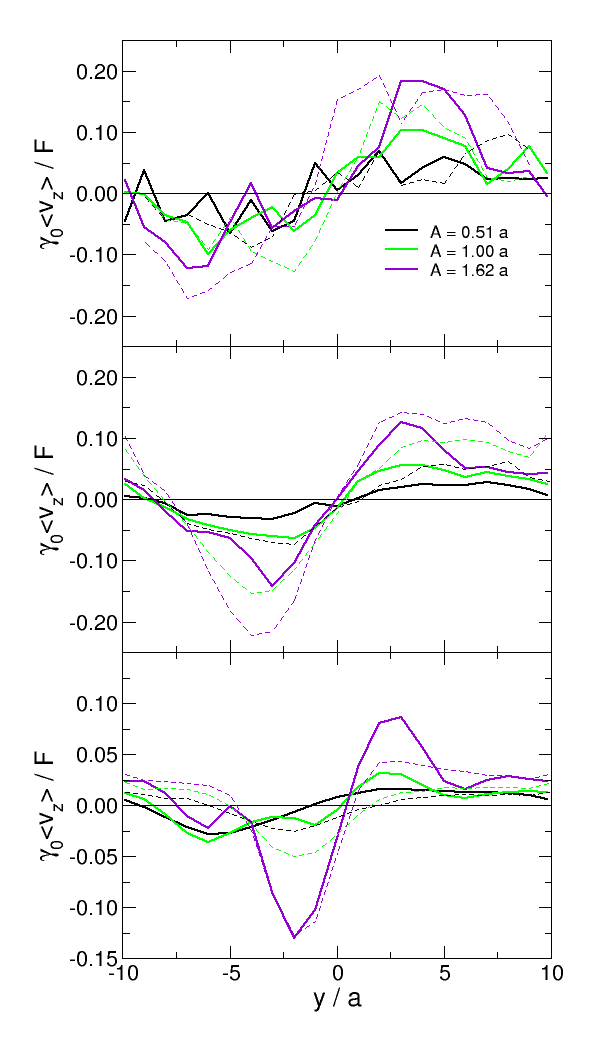}
\caption{Component of the tracer velocity along the transverse direction. 
The tracer velocity is averaged in slabs transverse to the channel  direction as a function of the channel corrugation and applied external force 
for different corrugations and forces: 
$\beta F_{ext}\lambda=10$ (upper panel), 
$\beta F_{ext}\lambda=100$ (intermediate panel) and 
$\beta F_{ext}\lambda=1000$ (bottom panel). 
Solid lines stand for the colloidal bath whereas dashed lines for the thermal bath.
\label{vz-A}}
\end{figure}
\clearpage
\newpage
\section*{First passage time distribution}
\begin{figure}[h!]
\includegraphics[scale=0.4]{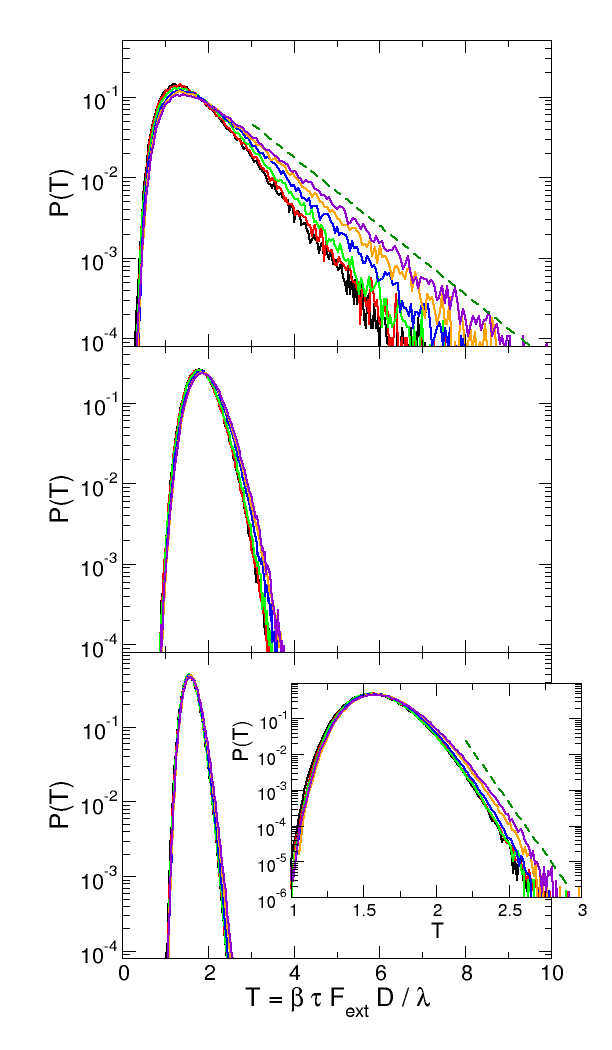}
\caption{Distributions of first passage times for different corrugations (as indicated in the legend) and forces: 
$\beta F_{ext}\lambda =10$ (upper panel), 
$\beta F_{ext}\lambda=100$ (intermediate panel) and 
$\beta F_{ext}\lambda=1000$ (bottom panel). The inset to the bottom panel shows a zoom to the proper scale. \label{passage-times}}
\end{figure}
\end{widetext}

\end{document}